\documentclass[12pt,a4paper]{article}
\usepackage[T2A]{fontenc}
\usepackage{amsfonts}
\usepackage{amsmath}
\usepackage{amssymb}
\usepackage[english]{babel}
\usepackage{graphicx}
\newtheorem{law1}{Conjecture}
\newtheorem{law2}{Theorem}

\begin{document}
\title{Discrete mechanics: a sequential growth dynamics for causal sets, and a self-organization of particles}
\author{Alexey L. Krugly\thanks{Quantum Information Laboratory, Institute of Physics and Physical Technologies, Moscow, Russia; akrugly@mail.ru.}}
\date{} \maketitle
\begin{abstract}
A model of a discrete pregeometry on a microscopic scale is introduced. This model is a finite network of finite elementary processes. The mathematical description is a d-graph that is a generalization of a graph. This is the particular case of a causal set. The aim of this study is to construct the particles as emergent structures. The particles in this model must be cyclic processes. The general dynamics and several examples are given. A simple dynamics generates a hierarchy of cyclic processes. An algebraic representation of this dynamics is given. It is based on the algebra of creation and destruction operators. Loops are described by bosonic operators and causal connections are described by fermionic operators.
\end{abstract}
\newpage
\tableofcontents
\newpage
\section{INTRODUCTION\label{I}}
Spacetime continuum is a macroscopic model. Einstein has considered clocks and rulers as a physical background of spacetime \cite{Einstein 1910}. Readouts on these devices fix points of spacetime. These points are physical events and have physically infinitesimal spatial sizes and durations. However in curved spacetime there can not be rigid rulers. Therefore other continuum of devices is used. These are standard pointlike clocks and pointlike light pulses. The standard of length is replaced with the standard of speed. In result all measurements are reduced to measurements of intervals of time. This method is called a chronometry \cite{Synge 1960}. Spacetime is the model of the properties of these chronometric measurements and nothing else. The reconstruction of spacetime is the reconstruction of the chronometric measurements.

Pseudo-Riemannian spacetime is a mathematical structure. It is based on some axioms. The axiomatic approach is well investigated \cite{Mould 1959,Vladimirov 1967,Pimenov 1968,Castagnino 1971,Enosh 1971,Ehlers 1972,Hawking 1976,Malament 1977}. This list of references is by no means complete. We know that the causal ordering of events in spacetime contains enough information to recover the topology, differential structure, and conformal metric. The rate of the standard pointlike clock unambiguously states the metric. These axioms are the idealized properties of the chronometric measurements. World lines of the standard pointlike clocks and the pointlike light pulses must be continuous. The rate of the standard pointlike clock depends on a gravitational field and does not depend on other fields. The emission and reception of the pointlike light pulses does not influence the rate of the standard pointlike clocks. The standard pointlike clock has infinitesimal weight and does not influence the gravitational field. Consequently, the standard pointlike clocks and the pointlike light pulses are classical devices. Spacetime continuum is a description of the properties of the macroscopic devices. This model is not valid on the microscopic scale. Quantum theory is not an adequate description of quantum objects. This is a description of measurements of quantum objects by classical devices. A self-consistent description of quantum objects can not use spacetime continuum.

One of approaches to quantum gravity is different models of a discrete pregeometry (see e.g. the introduction in \cite{0810.1768} for an attempt to address this issue). We expect quantum gravity to be described as a discrete theory of geometry, in some form or other. One can regard the continuum as emerging from the discrete, in similar way in which continuum theories for fluids emerge from the underlying physics of their discrete molecules. In the case of fluids, the discrete elements naturally live in a continuum background. The situation for gravity, however, is different. Discrete structure is viewed as fundamental, while continuum is emergent on a large scale only.

In the paper \cite{Krugly2010-1} I have introduced a model of a discrete pregeometry on a microscopic scale. I call it `a discrete mechanics'. This model is a particular case of a causal set. In this paper I consider a dynamics. The aim of this investigation is a self-consistent description of elementary particles without spacetime continuum. Some preliminary results are given in the paper \cite{Krugly2009-3}.

In the next section I introduce a review of the model. The next section contains all necessary information about the model because the paper \cite{Krugly2010-1} is written in Russian. In section \ref{SGD}, I consider the dynamics. In section \ref{ES}, I consider particular cases of the dynamics. There is a simple example of a self-organization. In section \ref{AR}, I introduce a algebraic formalism, and section \ref{CON} contains a short discussion.
\section{MODEL OF PREGEOMETRY}
\subsection{Elementary processes}
The first hypothesis of this study is that causality take place on the microscopic scale. In relativity theory the causal structure of the world is described by a partial order of events. In the instant of time there is a hypersurface of disconnected events. These events are connected only by intersections of the past light cones. Thus the structures of the physical world are not physical objects but physical processes.

The second hypothesis is a finite divisibility of any structures. Consequently, the central hypothesis of this study is that a physical process is a finite network of finite elementary processes \cite{STC,STC2,STC3,STC4}. A stable object is a stable process. This means a cyclic process. A scale hierarchy of the matter is a hierarchy of embedded cyclic processes.

The primitive entities of the physical world have not an internal structure. This is primordial indivisible objects. Consequently, they itself have not any internal properties except one. They exist. The property ``existence'' can adopt two values: ``the primitive entity exists'', and ``the primitive entity does not exist''. The primitive entity is called a material point. The primitive process can be thought of as act of creation. The value of the property ``existence'' of the material point varies from ``the material point does not exist'' to ``the material point exists'' by this process. Its dual represents the act of destruction. These primitive processes are called monads \cite{STC4}. A propagation of the material point is simply an ordered pair of creation and annihilation. This process of propagation is called a chronon \cite{STC}. The general process will be a collection of creations and annihilations. We have a following conjecture.

\begin{law1}\label{P01} Every physical process is a finite combination of finite elementary processes, monads. There are two kinds of monads: the creation and the annihilation of a primitive entity or a material point. A primitive entity itself has not any internal structure and any internal properties except "existence". An ordered pair of creation and annihilation is the propagation of a primitive entity, or a chronon.
\end{law1}

A physical theory has to do with physical facts and with mathematical or linguistic symbols for them. We can express the fundamental relations as algebraic relations. We have a two-state material point: the state $|1\rangle$ means the material point exists, while the state $|0\rangle$ means the material point does not exist. We denote its creation and destruction operators by $\hat a^{\dag}$ and $\hat a$, respectively. By definition, put
\begin{equation}
\label{eq:mp1.1}
|1\rangle =\hat a^{\dag}|0\rangle \textrm{,}
\end{equation}
\begin{equation}
\label{eq:mp1.01}
|0\rangle =\hat a|0\rangle \textrm{.}
\end{equation}
This is two elementary processes or monads: a creation and a destruction. The algebraic representation will be used in section \ref{AR}.

\begin{figure}
	\centering	
		\includegraphics[trim=8cm 18cm 8cm 9cm]{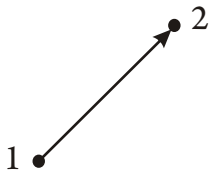}
	\caption{A chronon.}
	\label{fig:fig1}
\end{figure}

We can describe a chronon in terms of graph theory. It is a directed edge (Fig.\ \ref{fig:fig1}). The vertex 1 is a creation, the vertex 2 is a destruction. The direction of the edge expresses an immediate causal priority of the monads.

\begin{figure}
	\centering	
		\includegraphics[trim=8cm 18cm 8cm 7cm]{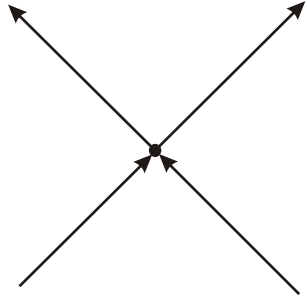}
	\caption{A x-structure.}
	\label{fig:fig2}
\end{figure}

Suppose a self-action is impossible. The material point can be destroyed only by interaction with another material point. The interaction of this second material point means the change of its state. Only one kind of change is possible. This is the annihilation of the second material point. Suppose the number of the material points does not change. This is a fundamental conservation law. We have a simplest interaction process: two material points are destroyed and two material points are created. This process is called a tetrad \cite{FinkMcC1} or an x-structure \cite{Krugly2002} (Fig.\ \ref{fig:fig2}). The x-structure consists of two monads of destruction and two monads of creation. Suppose any process can be divided into x-structures. This symmetric dyadic kinematics is called for brevity X kinematics \cite{Fink88}. We have a following conjecture.

\begin{law1}\label{P02} There is one kind of interactions: two material points are destroyed and two material points are created. This process is called a x-structure. Any process consists of x-structures.
\end{law1}

A chronon and an x-structure describe an immediate causal priority of monads. If a creation causally precedes an annihilation this is a chronon. If an annihilation causally precedes a creation this is an x-structure. Suppose there is an universal causal order of monads. We have a following conjecture.

\begin{law1}\label{P03} Any process is a causal set of monads. A causal relation and an immediate causal priority are consistent.
\end{law1}

A causal set is a set $\mathcal{C}$ endowed with a binary relation `precedes', $\prec$, that satisfies for any elements $a,\ b,\ c$:
\begin{equation}
\label{eq:mp1.2} a\prec a\qquad\textrm{(irreflexivity),}
\end{equation}
\begin{equation}
\label{eq:mp1.3} \{a\mid(a\prec b)\wedge(b\prec a)\}=\emptyset \qquad \textrm{(acyclicity),}
\end{equation}
\begin{equation}
\label{eq:mp1.4} (a\prec b)\wedge(b\prec c)\Rightarrow(a\prec c)\qquad \textrm{(transitivity),}
\end{equation}
\begin{equation}
\label{eq:mp1.5} \mid\mathcal{A}(a, b)\mid<\infty\qquad\textrm{(local finiteness),}
\end{equation}
where $\mathcal{A}(a,\ b)$ is an Alexandrov set of the elements $a$ è $b$. $\mathcal{A}(a,\ b)=\{c\mid a\prec c \prec b\}$. The local finiteness means that the Alexandrov set of any elements is finite. Sets of monads are denoted by calligraphic capital Latin letters. Causal set hypothesis is introduced in \cite{Myrheim,'t Hooft}. For comprehensive reviews of the field see, for example, \cite{Sorkin2005,Dowker2006,Henson2009,Wallden2010}.
\subsection{A d-graph}
In the considered model any process can be described as some graph. But such description is not useful. By definition, a graph is a set of vertexes and a binary relation (edges) over this set. We cannot describe external lines as in Feynman diagrams. For example, an x-structure is not a graph. We can define a set of edges, and vertexes as a relation over this set. But in this case, if we divide a structure into substructures we must duplicate the edges which connect these substructures. This is not useful either. It is convenient to break the edge into two halves, monads, of which the edge is regarded as composed \cite{FinkMcC1}.

Introduce the axiomatic approach to this model. Consider the set $\mathcal{G}$ of monads and a binary relation (an immediate causal priority) over this set. By $\alpha_i$ and $\beta_j$ denote the monad of creation and destruction, respectively. By $(\alpha_i\beta_j)$ denote an immediate causal priority relation of $\alpha_i$ and $\beta_j$. $\mathcal{G}$ satisfies the following axioms.
\begin{equation}
\label{eq:mp1.6}
\forall\alpha_i(\exists !\beta_j(\alpha_i\beta_j))\vee(\not\exists\beta_j(\alpha_i\beta_j))\textrm{,}
\end{equation}
\begin{equation}
\label{eq:mp1.7}
\forall\alpha_i(\not\exists \alpha_j(\alpha_i\alpha_j))\textrm{,}
\end{equation}
\begin{equation}
\label{eq:mp1.8}
\forall\beta_j(\exists !\alpha_i(\alpha_i\beta_j))\vee(\not\exists\alpha_i(\alpha_i\beta_j))\textrm{,}
\end{equation}
\begin{equation}
\label{eq:mp1.9}
\forall\beta_j(\not\exists \beta_i(\beta_i\beta_j))\textrm{.}
\end{equation}
There is no more than one monad $\alpha_i$ and does not exist the monad $\beta_i$ which immediately causally precede any $\beta_j$. There is no more than one monad $\beta_j$ and does not exist the monad $\alpha _j$ which immediately causally follow $\alpha_i$. The pair $(\alpha_i\beta_j)$ is called a chronon or an edge.

The following axioms describe an x-structure.
\begin{equation}
\label{eq:mp1.10}
\forall\alpha_i\exists !\alpha_j(\forall\beta_k(\beta_k\alpha_i)\Rightarrow(\beta_k\alpha_j))\textrm{,}
\end{equation}
\begin{equation}
\label{eq:mp1.11}
\forall\beta_i\exists !\beta_j(\forall\alpha_k(\beta_i\alpha_k)\Rightarrow(\beta_j\alpha_k))\textrm{.}
\end{equation}
There is two and only two monads $\beta_i$ and $\beta_j$ which immediately causally precede any $\alpha_k$. There is two and only two monads $\alpha_i$ and $\alpha_j$ which immediately causally follow any $\beta_k$.

A causality is described by the following axiom.
\begin{equation}
\label{eq:mp1.12}
\{\alpha_i|(\alpha_i\beta_k)(\beta_k\alpha_l)\dots(\beta_j\alpha_i)\}= \emptyset\textrm{.}
\end{equation}
We do not discuss the problem of the finiteness or infinity of the universe. In any case an observer can have only a finite amount of information. Then we consider only finite sets of monads.
\begin{equation}
\label{eq:mp1.13} |\mathcal{G}|<\infty\textrm{.}
\end{equation}

We consider $\mathcal{G}$ for the description of a dynamics of pregeometry. Let call $\mathcal{G}$ a dynamical graph or a d-graph. Any $\mathcal{G}$ is a d-subgraph of the d-graph of the universe.

Some definitions and properties will be useful for the following work. The proofs are in \cite{Krugly2010-1}.

The monad of any type is denoted by $\gamma_i$. The monad $\gamma_i$ may be $\alpha_i$ or $\beta_i$.

Any monad belongs to an x-structure. Any d-graph can be formed from the empty set of monads by sequential adding of x-structures one after another.

\begin{figure}
	\centering	
		\includegraphics[width=4cm,trim=8cm 4cm 8cm 9cm]{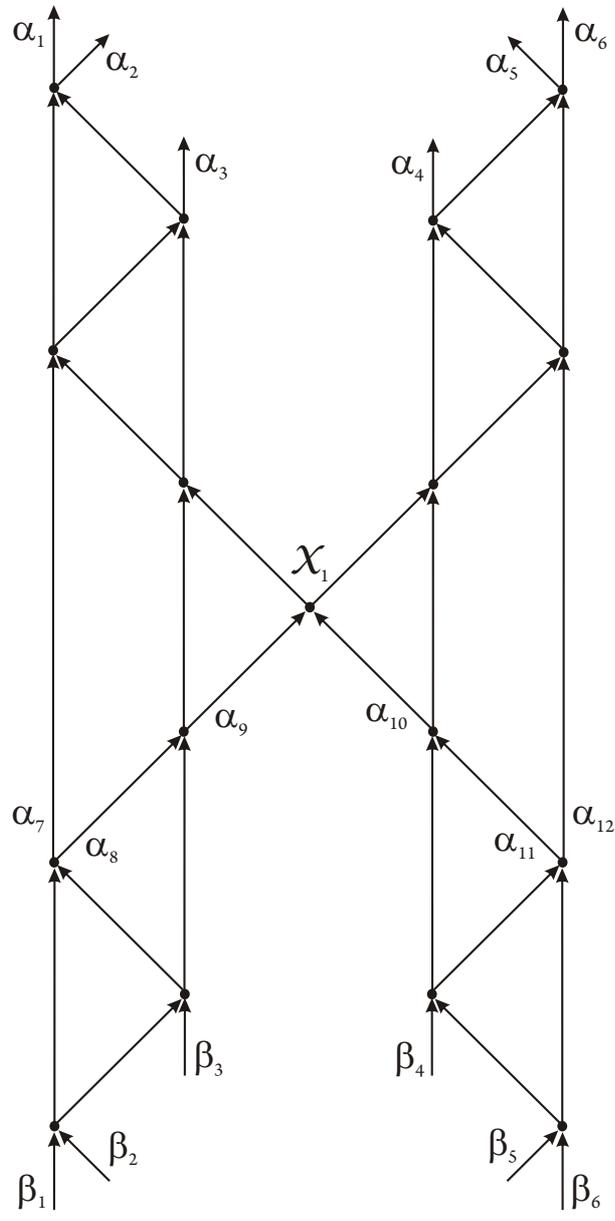}
	\caption{The example of a d-graph.}
	\label{fig:fig3}
\end{figure}

Define the isomorphism that takes each x-structure to the vertex of some graph and each chronon to the edge of this graph. We get the directed acyclic graph. All properties of this graph are the properties of the d-graph. For this reason I have called the considered set of monads in this model a d-graph. This isomorphism is used in the figures for simplicity. The edges are figured without a partition into monads. The monads are figured by placing $\alpha_i$ and $\beta_j$ near the beginning and the end of the edge, respectively, as necessary. The x-structures are figured by big black points. The example can be seen in Fig.\ \ref{fig:fig3}.

A sequence of monads is a saturated chain or a path if each monad immediately causally precedes the sequent monad.

Two monads are causally connected if they are connected by the path. The causal connection is denoted by $\prec$. The first monad of the path is called a cause. The last monad of the path is called an effect. Two monads $\gamma_i$ and $\gamma_j$ are causally unconnected iff neither $\gamma_i\prec\gamma_j$ nor $\gamma_j\prec\gamma_i$.

By definition, put $\mathcal{A}(\gamma_i,\ \gamma_j)=\{\gamma_s| \gamma_i\prec \gamma_s\prec \gamma_i\}$. The set $\mathcal{A}(\gamma_i,\ \gamma_j)$ is called an Alexandrov set of $\gamma_i $ and $\gamma_j$. By definition, put $\mathcal{\tilde A}(\gamma_i,\ \gamma_j)=\{\gamma_s|\gamma_i\preceq \gamma_s\preceq \gamma_i\}$. The set $\mathcal{\tilde A}(\gamma_i,\ \gamma_j)$ is called an inclusive Alexandrov set of $\gamma_i $ and $\gamma_j$. Consider two x-structures $\mathcal{X}_i=\{\alpha_{i1}, \alpha_{i2}, \beta_{i1}, \beta_{i2}\}$ and $\mathcal{X}_j=\{\alpha_{j1}, \alpha_{j2}, \beta_{j1}, \beta_{j2}\}$. Let $\mathcal{A}(\mathcal{X}_i,\ \mathcal{X}_j)$ be an Alexandrov set of the x-structure $\mathcal{X}_i$ and $\mathcal{X}_j$ if $\mathcal{A}(\mathcal{X}_i,\ \mathcal{X}_j)= \mathcal{A}(\beta_{i1}, \alpha_{j1})=\mathcal{A}(\beta_{i2}, \alpha_{j1}) =\mathcal{A}(\beta_{i1}, \alpha_{j2})=\mathcal{A}(\beta_{i2}, \alpha_{j2})$.

Two monads are related by the immediate causal priority if and only if they are causally connected and them Alexandrov set is empty.

A d-graph is a causal set of monads, and a causal set of chronons, and a causal set of x-structures. The subset of monads $\alpha_i$ and the subset of monads $\beta_j$ are causal sets.

The past of the monad is the set of monads, which causally precede this monad. The past of $\gamma_i$ is denoted by $\mathcal{P}(\gamma_i)=\{\gamma_j|\gamma_j\prec\gamma_i\}$. The future of the monad is the set of monads, which causally follow this monad. The future of $\gamma_i$ is denoted by $\mathcal{F}(\gamma_i)=\{\gamma_j|\gamma_j\succ\gamma_i\}$. Consider x-structure $\mathcal{X}_i=\{\alpha_{i1}, \alpha_{i2}, \beta_{i1}, \beta_{i2}\}$. By definition, put $\mathcal{P}(\mathcal{X}_i)= \mathcal{P}(\alpha_{i1})= \mathcal{P}(\alpha_{i2})$ and $\mathcal{F}(\mathcal{X}_i)= \mathcal{F}(\beta_{i1})= \mathcal{F}(\beta_{i2})$.

A monad is called maximal iff its future is an empty set. Any maximal monad is a monad of a type $\alpha$. In Fig.\ \ref{fig:fig3} $\alpha_1$, $\alpha_2$, $\alpha_3$, $\alpha_4$, $\alpha_5$, and $\alpha_6$ are maximal monads. A monad is called minimal iff its past is an empty set. Any minimal monad is a monad of a type $\beta$. In Fig.\ \ref{fig:fig3} $\beta_1$, $\beta_2$, $\beta_3$, $\beta_4$, $\beta_5$, and è $\beta_6$ are minimal monads. Maximal and minimal monads are called external monads. Other monads are called internal monads. The monad is internal iff it is included in a chronon.

A chain is a totally (or a linearly) ordered subset of monads. Every two monads of this subset are related by $\prec$. A chain is a subset of a path.

An antichain is a totally unordered subset of monads. Every two elements of this subset are not related by $\prec$. The cardinality of an antichain is called a width of an antichain.

A slice is a maximal antichain. The maximal antichain cannot be enlarged and remain an antichain. Equivalently, every monad in $\mathcal{G}$ is either in the slice or causal connected to one of its monads. In Fig.\ \ref{fig:fig3} the set $\{\alpha_7\ \alpha_8\ \alpha_9\ \alpha_{10}\ \alpha_{11}\ \alpha_{12}\}$ is the example of a slice. The set of all maximal (or minimal) monads is a slice. A slice is a discrete spacelike hypersurface.

All slices of $\mathcal{G}$ have the same width. This width is called a width of $\mathcal{G}$. In Fig.\ \ref{fig:fig3} the width of the d-graph is equal to 6. This is the conservation law of the number of material points. The number of material points remains in each elementary interaction (x-structure). Consequently the number of material points remains in each process. We can consider the d-graph in Fig.\ \ref{fig:fig3} as two processes. The width of each process is equal to 3. These processes interact by the x-structure $\mathcal{X}_1$.
\section{SEQUENTIAL GROWTH DYNAMICS\label{SGD}}
\subsection{Principles of dynamics\label{PD}}
Consider the following concept of a d-graph dynamics. The past and the future exist, are determined, and are changeless. This concept is opposite to the concept of an emergent future. For example, these two concepts are described in the introduction of \cite{HR1956}. It is possible, we can interpret the considered dynamics in the concept of emergent future but we do not discuss this problem. In the discrete mechanics this means that the d-graph of the universe exists. It is meaningless to talk about the exact structure of the d-graph if we cannot determine its structure. The structure of the d-graph of the universe implies the infinite amount of information. But we can only actually know a finite number of facts. Therefore any observer can consider only finite fragments and take into account the rest of the d-graph of the universe in an approximate way. We have the following assumption.

\begin{law1}\label{P31} Any d-graph has the certain structure. We can determine the structure of any d-graph.
\end{law1}

We can consider this conjecture as a consequence of a causality principle. The causality means the certain causal order. This order is the structure of the d-graph.

Suppose we have the information about the structure of some d-graph $\mathcal{G}$. This is the description of some part of some physical process. The task is to predict the future stages of this process or to reconstruct the past stages. This mean to determine the structure of the d-subgraph that is connected with $\mathcal{G}$. In general case, we cannot determine this structure unambiguously. We can only calculate probabilities of different variants. In particular case, the probability of some variant can be equal to 1. This is a deterministic process.

The aim of the d-graph dynamics is to calculate probabilities of the structures of any d-graphs $\mathcal{G}$ that is connected with the given $\mathcal{G}_0$. We can reconstruct the structure of $\mathcal{G}$ step by step. This procedure is called the sequential growth dynamics \cite{RideoutSorkin}. The growth of the connected d-graph is a sequence of some elementary processes. We sequentially add new parts to $\mathcal{G}_0$. The minimal part is an x-structure. The growth is the addition of new x-structures one after another \cite{ Krugly2002,Krugly1998}. The addition of one x-structure is called an elementary extension.

We can determine the structure of the d-graph after each elementary extension by conjecture \ref{P31}. This is not an appearance of new parts of the d-graph of the universe. This is an appearance of new information about the existing d-graph of the universe. We can randomly initiate one elementary extension for any given d-graph and we can determine the exact change of the structure of the d-graph that is the result of this elementary extension. This procedure is called the elementary measurement. The sequence of elementary extensions is the sequence of the obtaining of the information about the structure of the d-graph by the observer.

A set of results of sequential measurements is a classical stochastic sequence. Thus the sequence of the elementary extensions is a classical stochastic sequence.

In this concept of the dynamics we assume that the complex amplitudes in quantum theory are not fundamental quantities \cite{Krugly2009-1,Krugly2009-2}. This is a mathematical tool for the approximate description of d-graphs by the matter in continuous spacetime.

Consider all types of elementary extensions.

First type is an internal elementary extension to the future (Fig.\ \ref{fig:fig4}). In this and following figures the d-graph $\mathcal{G}$ is represented by a rectangle because it can have an arbitrary structure. This extension describes the future evolution of the process without any interaction with an environment. The width $n$ of $\mathcal{G}$ is not changed by this elementary extension.
\begin{figure}
	\centering	
		\includegraphics[width=4cm,trim=8cm 16cm 8cm 5cm]{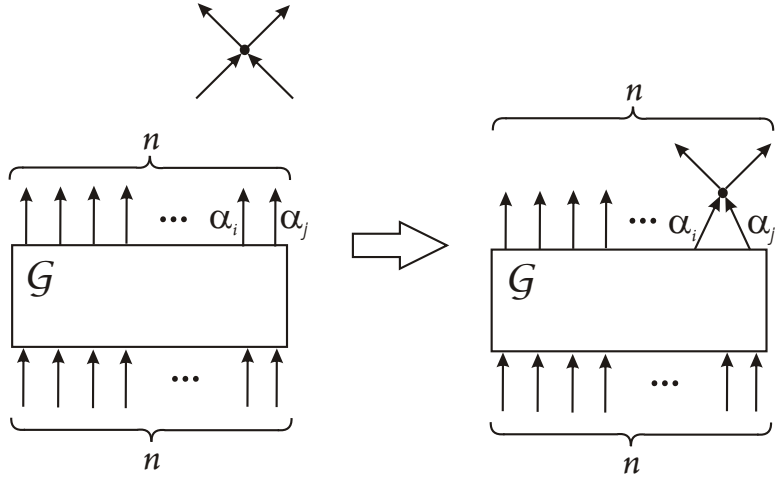}
	\caption{The internal elementary extension to the future.}
	\label{fig:fig4}
\end{figure}

\begin{figure}
	\centering	
		\includegraphics[width=4cm,trim=8cm 16cm 8cm 5cm]{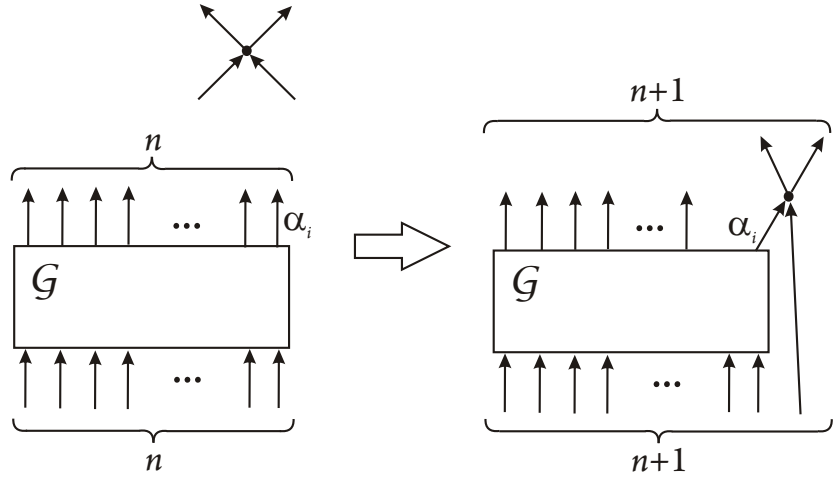}
	\caption{The external elementary extension to the future.}
	\label{fig:fig5}
\end{figure}

Second type is an external elementary extension to the future (Fig.\ \ref{fig:fig5}). This extension describes the future interaction of the process with an environment. The width $n$ of $\mathcal{G}$ has increased by 1.

\begin{figure}
	\centering	
		\includegraphics[width=4cm,trim=8cm 14cm 8cm 7cm]{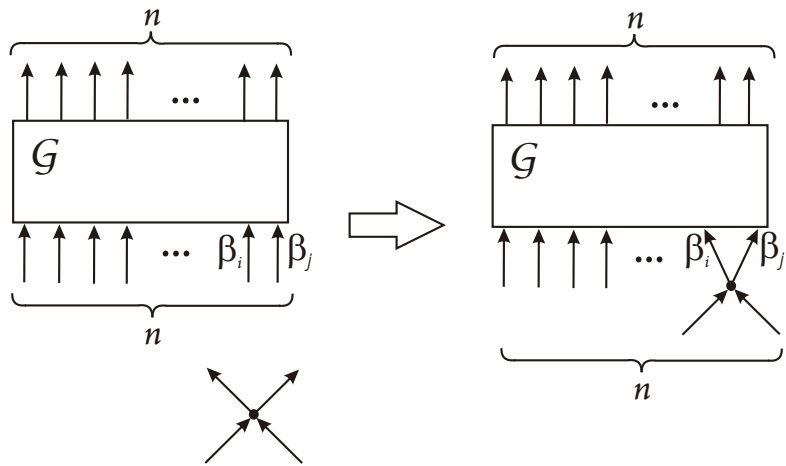}
	\caption{The internal elementary extension to the past.}
	\label{fig:fig6}
\end{figure}

Third type is an internal elementary extension to the past (Fig.\ \ref{fig:fig6}). This extension describes the past evolution of the process without any interaction with an environment. The width $n$ of $\mathcal{G}$ is not changed by this elementary extension.

\begin{figure}
	\centering	
		\includegraphics[width=4cm,trim=8cm 14cm 8cm 7cm]{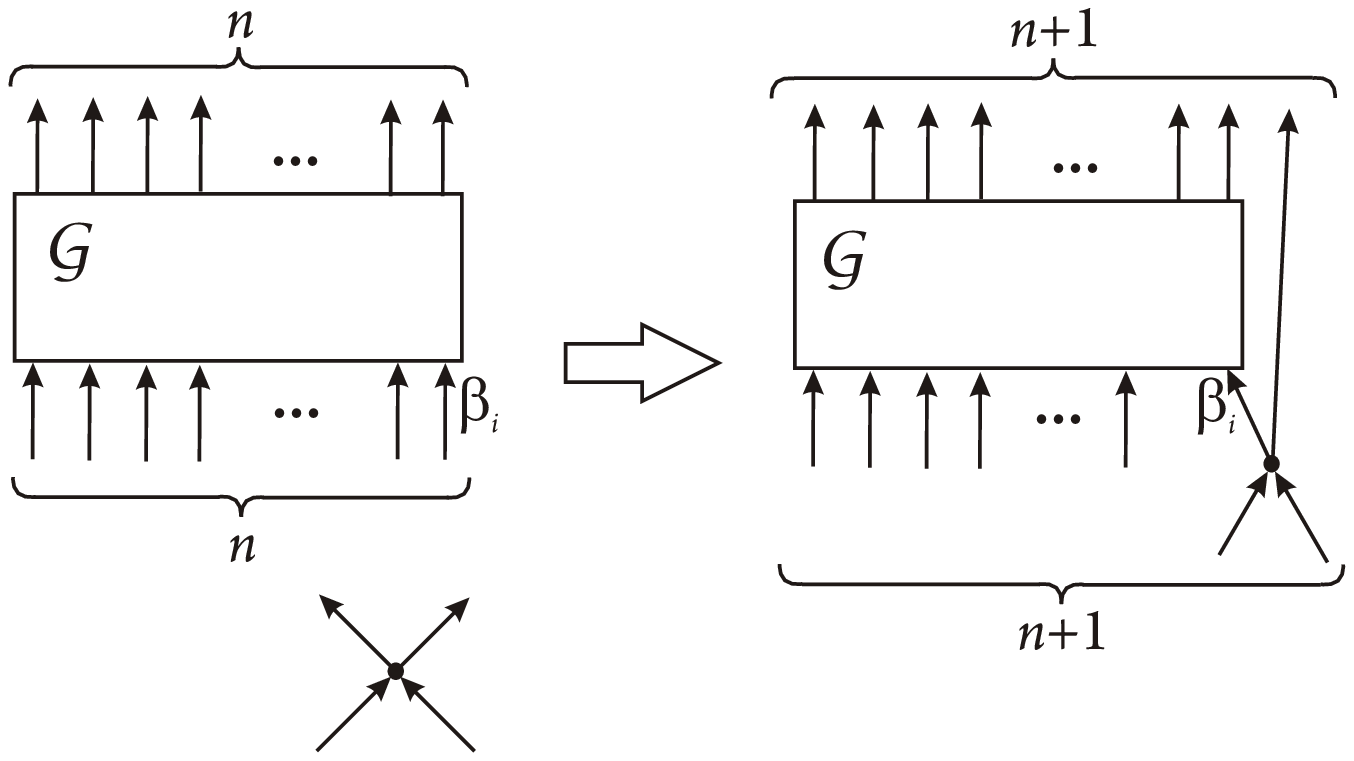}
	\caption{The external elementary extension to the past.}
	\label{fig:fig7}
\end{figure}

Fourth type is an external elementary extension to the past (Fig.\ \ref{fig:fig7}). This extension describes the past interaction of the process with an environment. The width $n$ of $\mathcal{G}$ has increased by 1.

\begin{figure}
	\centering	
		\includegraphics[width=4cm,trim=8cm 13cm 8cm 7cm]{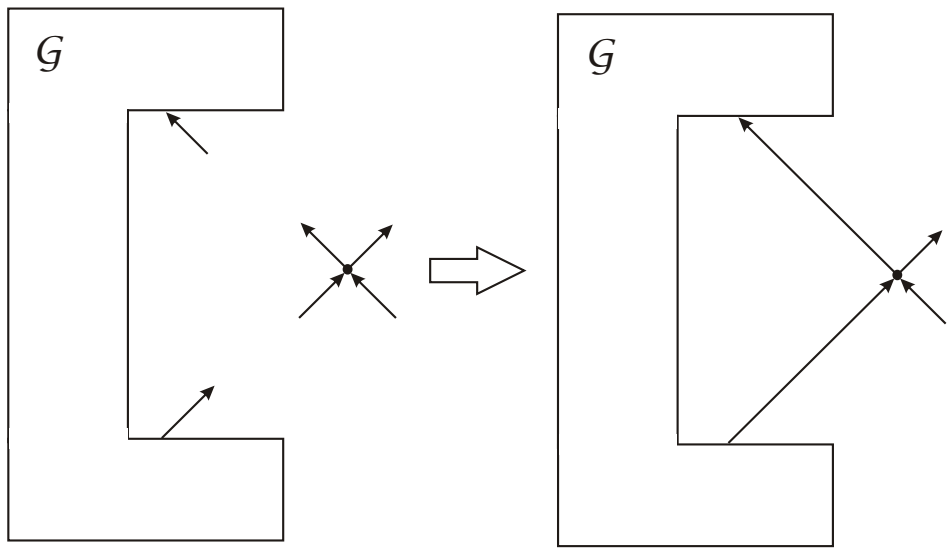}
	\caption{The forbidden elementary extension.}
	\label{fig:fig8}
\end{figure}

We can get any connected d-graph by the elementary extensions of these types. Other elementary extensions are unnecessary. We consider the simplest variant of the sequential growth dynamics. We suppose that the law of this dynamics has a simplest form. Similarly the laws of classical mechanics have a simplest form in an inertial frame of reference. We forbid another type of the elementary extensions (Fig.\ \ref{fig:fig8}). This elementary extension puts the x-structure into the d-graph.

If we can calculate the probabilities of all elementary extensions of any d-graph, we can calculate the probabilities of all new parts of any d-graph as the probabilities of random sequences of elementary extensions.
\subsection{The normalization of the probabilities \label{N}}
Let's number all possible elementary extensions of $\mathcal{G}$. Let $P_{a}$ be the probability of the elementary extension number $a$. All information consists in the structure of the d-graph. We have
\begin{equation}
\label{eq:sgd2.1} P_{a}=Cp_a\textrm{,}
\end{equation}
where $p_a$ is the function of the structure of the $\mathcal{G}$, and $C$ is the normalization constant. In general case, $p_a$ and $C$ depend on the type of the elementary extension.

The internal elementary extension to the future (Fig.\ \ref{fig:fig4}) is the addition of new x-structure to two maximal monads $\alpha_i$ and $\alpha_j$ of $\mathcal{G}$. Indices $i$ and $j$ range from 0 to $n$, where $n$ is the width of $\mathcal{G}$. We have $(1/2)n(n-1)$ such elementary extensions. Let $P_{ij(int,f)}$ be the probability of this elementary extension. Consider the sum $P_{int,f}$ of the probabilities of all internal elementary extensions to the future. We have
\begin{equation}
\label{eq:sgd2.2} P_{int,f}=\sum_{i=1}^{n}\sum_{j>i}^{n}P_{ij(int,f)}\textrm{.}
\end{equation}

The external elementary extension to the future (Fig.\ \ref{fig:fig5}) is the addition of new x-structure to one maximal monad $\alpha_i$ of $\mathcal{G}$. We have $n$ such elementary extensions. Let $P_{i(ext,f)}$ be the probability of this elementary extension. Consider the sum $P_{ext,f}$ of the probabilities of all external elementary extensions to the future. We have
\begin{equation}
\label{eq:sgd2.3} P_{ext,f}=\sum_{i=1}^{n}P_{i(ext,f)}\textrm{.}
\end{equation}

Similarly, the internal elementary extension to the past (Fig.\ \ref{fig:fig6}) is the addition of new x-structure to two minimal monads $\beta_i$ and $\beta_j$ of $\mathcal{G}$. Indices $i$ and $j$ range from 0 to $n$, where $n$ is the width of $\mathcal{G}$. The external elementary extension to the past (Fig.\ \ref{fig:fig7}) is the addition of new x-structure to one minimal monad $\beta_i$ of $\mathcal{G}$. Consider the sum $P_{int,p}$ of the probabilities $P_{ij(int,p)}$ of all internal elementary extensions to the past and the sum $P_{ext,p}$ of the probabilities $P_{i(ext,p)}$ of all external elementary extensions to the past. We have
\begin{equation}
\label{eq:sgd2.4} P_{int,p}=\sum_{i=1}^{n}\sum_{j>i}^{n}P_{ij(int,p)}\textrm{,}
\end{equation}
\begin{equation}
\label{eq:sgd2.5} P_{ext,p}=\sum_{i=1}^{n}P_{i(ext,p)}\textrm{.}
\end{equation}

We must require that the sum of the full set of the probabilities of the elementary extensions issuing from a given d-graph be unity.
\begin{equation}
\label{eq:sgd2.6} P_{int,f}+ P_{ext,f}+ P_{int,p}+ P_{ext,p}=1\textrm{.}
\end{equation}

By definition, put
\begin{equation}
\label{eq:sgd2.7} K_t=\frac{ P_{int,f}+ P_{ext,f}}{P_{int,p}+ P_{ext,p}}\textrm{,}
\end{equation}
\begin{equation}
\label{eq:sgd2.8} K_{int,f}= P_{int,f}/ P_{ext,f}\textrm{,}
\end{equation}
\begin{equation}
\label{eq:sgd2.9} K_{int,p}= P_{int,p}/ P_{ext,p}\textrm{.}
\end{equation}
The coefficient $K_t$ describes the time asymmetry of the dynamics. The coefficient $K_{int,f}$ describes the intensity of the interaction between $\mathcal{G}$ and the environment for the future evolution. The coefficient $K_{int,f}$ describes the intensity of the interaction between $\mathcal{G}$ and the environment for the past evolution. We can choose any values of these coefficients. In following consideration we assume the time symmetry:
\begin{equation}
\label{eq:sgd2.90} K_t=1\textrm{,}
\end{equation}
\begin{equation}
\label{eq:sgd2.91} K:=K_{int,f}=K_{int,p}\textrm{.}
\end{equation}

If we choose the coefficients (\ref{eq:sgd2.7}) - (\ref{eq:sgd2.9}) we unambiguously choose the probabilities $P_{int,f}$, $P_{ext,f}$, $P_{int,p}$, and $P_{ext,p}$. This determines the normalization constants. We have
\begin{equation}
\label{eq:sgd2.10} C_{int,f}=\frac{P_{int,f}}{\sum_{i=1}^{n}\sum_{j>i}^{n} p_{ij(int,f)}}\textrm{,}
\end{equation}
where $p_{ij(int,f)}$ is the function of the structure of the $\mathcal{G}$, and $C_{int,f}$ is the normalization constant for the internal elementary extensions to the future.
\begin{equation}
\label{eq:sgd2.11} C_{ext,f}=\frac{P_{ext,f}}{\sum_{i=1}^{n}p_{i(ext,f)}}\textrm{,}
\end{equation}
where $p_{i(ext,f)}$ is the function of the structure of the $\mathcal{G}$, and $C_{ext,f}$ is the normalization constant for the external elementary extensions to the future.
\begin{equation}
\label{eq:sgd2.12} C_{int,p}=\frac{P_{int,p}}{\sum_{i=1}^{n}\sum_{j>i}^{n} p_{ij(int,p)}}\textrm{,}
\end{equation}
where $p_{ij(int,p)}$ is the function of the structure of the $\mathcal{G}$, and $C_{int,p}$ is the normalization constant for the internal elementary extensions to the past.
\begin{equation}
\label{eq:sgd2.13} C_{ext,p}=\frac{P_{ext,p}}{\sum_{i=1}^{n}p_{i(ext,p)}}\textrm{,}
\end{equation}
where $p_{i(ext,p)}$ is the function of the structure of the $\mathcal{G}$, and $C_{ext,p}$ is the normalization constant for the external elementary extensions to the past.
\subsection{The dynamical causality\label{DC}}
The function $p_a$ in (\ref{eq:sgd2.1}) can depend only on the structure of some d-subgraph $\mathcal{G}_a$ of $\mathcal{G}$. $\mathcal{G}_a$ is called a cause of the considered elementary extension number $a$ if this is the elementary extension to the future. Otherwise if this is the elementary extension to the past, $\mathcal{G}_a$ is called an effect of the considered elementary extension. The physical idea is that this dynamical causality is consistent with causal order of monads. In this way, the order relation of a d-graph will be causal in the dynamical sense, and not only in name \cite{RideoutSorkin}.

Consider the elementary extension to the past number $a$. This is the addition of new x-structure $\mathcal{X}_a=\{\alpha_{a1}, \alpha_{a2}, \beta_{a1}, \beta_{a2}\}$ to $\mathcal{G}_0$. We get $\mathcal{G}_0\cup\mathcal{X}_a$. Also consider the elementary extension to the future number $b$. This is the addition of new x-structure $\mathcal{X}_b=\{\alpha_{b1}, \alpha_{b2}, \beta_{b1}, \beta_{b2}\}$ to $\mathcal{G}_0$. We get $\mathcal{G}_0\cup\mathcal{X}_b$. We assume the following dynamical causality.

\begin{law1}\label{P32} The function $p_a$ can depend only on the new x-structure $\mathcal{X}_a$ and on the future of $\mathcal{X}_a$ for the elementary extension to the past number $a$.
\begin{equation}
\label{eq:sgd2.14} p_a= p_a(\mathcal{F}(\mathcal{X}_a)\cup \mathcal{X}_a)\textrm{,}
\end{equation}
where $\mathcal{F}(\mathcal{X}_a)\in\mathcal{G}_0$.

The function $p_b$ can depend only on the new x-structure $\mathcal{X}_b$ and on the past of $\mathcal{X}_b$ for the elementary extension to the future number $b$.
\begin{equation}
\label{eq:sgd2.15} p_b= p_b(\mathcal{P}(\mathcal{X}_b)\cup \mathcal{X}_b)\textrm{,}
\end{equation}
where $\mathcal{P}(\mathcal{X}_b)\in\mathcal{G}_0$.
\end{law1}

\begin{figure}
	\centering	
		\includegraphics[width=4cm,trim=8cm 16cm 8cm 6cm]{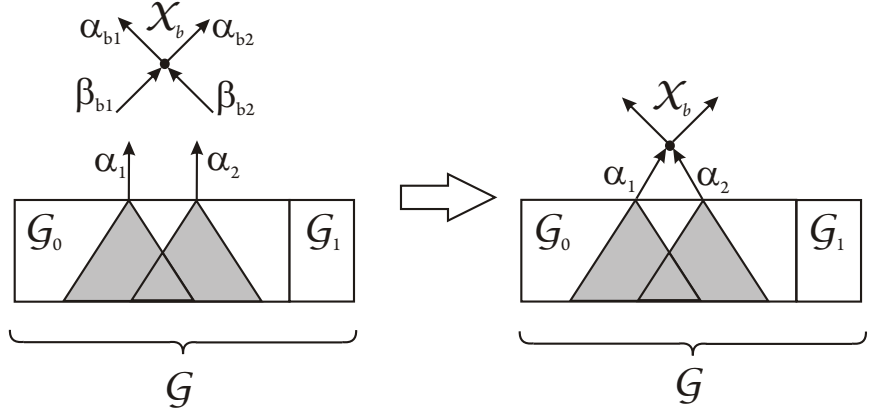}
	\caption{The past of the elementary extension (the shaded region).}
	\label{fig:fig9}
\end{figure}

The normalization constant can depend on another part of $\mathcal{G}_0$. This differs from the dynamical causality in \cite{RideoutSorkin}. Consider the example. This is the internal elementary extension to the future of $\mathcal{G}_0$ that is the addition of x-structure $\mathcal{X}_b$ to the maximal monads $\alpha_1$ and $\alpha_2$ (Fig.\ \ref{fig:fig9}). 
We have (\ref{eq:sgd2.1}) for the probability of this elementary extension. By assumption, $p_b$ is a function of $\mathcal{P}(\mathcal{X}_b)\cup\alpha_{b1}\cup\alpha_{b2}$. But in general case, $C$ depends on the whole $\mathcal{G}_0$. Add the d-subgraph $\mathcal{G}_1$ to $\mathcal{G}_0$ such that $(\mathcal{P}(\alpha_1)\cup\mathcal{P}(\alpha_2)\cup\alpha_1\cup\alpha_2)\cap\mathcal{G}_1=\emptyset$. We get $\mathcal{G}=\mathcal{G}_0\cup\mathcal{G}_1$. Consider the internal elementary extension to the future of $\mathcal{G}$ that is the addition of x-structure $\mathcal{X}_b$ to the maximal monads $\alpha_1$ and $\alpha_2$. The function $p_b$ is the same. But the normalization constants are different. We have (\ref{eq:sgd2.10}). This sum includes the internal elementary extensions to the future of $\mathcal{G}_1$ in the case of $\mathcal{G}$ and does not include they in the case of $\mathcal{G}_0$.
\subsection{The discrete general covariance.\label{DGC}}
We can construct any d-graph from the empty set by the sequence of the elementary extensions. First x-structure appears (say with probability one, since the universe exists). In general case, we can construct the d-graph from the x-structure by different sequences of the elementary extensions. Let $P_{ad}$ be the probability of the elementary extension number $a$ in the sequence number $d$. If $\mathcal{G}$ consists of $N$ x-structures we can obtain it from the empty set by the sequence of $N$ elementary extensions. Let $P(\mathcal{G})$ be the total probability of $\mathcal{G}$. We have
\begin{equation}
\label{eq:sgd4.1} P(\mathcal{G})=\sum_{d}\prod_{a=1}^{N}P_{ad} \textrm{,}
\end{equation}
where $d$ ranges over all sequences of $N$ elementary extensions that generate $\mathcal{G}$. The numbering of monads of $\mathcal{G}$ carries no physical meaning. We consider the two d-graphs as the same d-graph if they differ only in numbering of monads. Consequently the summation (\ref{eq:sgd4.1}) includes all sequences of $N$ elementary extensions that generate $\mathcal{G}$ with the different numbering of monads.

The probability of the sequence number $d$ of $N$ elementary extensions can be written as
\begin{equation}
\label{eq:sgd4.2} \prod_{a=1}^{N}P_{ad}= \prod_{a=1}^{N}C_{ad}p_{ad} \textrm{,}
\end{equation}
where $ p_{ad}$ is the function of the structure of $\mathcal{G}$ and $C_{ad}$ is the the normalization constant for the elementary extension number $a$.

We can start from any d-graph $\mathcal{G}_0$ instead of empty set. Let $\mathcal{G}_0$ consists of $N_0$ x-structures. If we can get $\mathcal{G}$ by adding $N-N_0$ x-structures to $\mathcal{G}_0$ we can consider the conditional probability of the transition from $\mathcal{G}_0$ to $\mathcal{G}$. We have
\begin{equation}
\label{eq:sgd4.3} P(\mathcal{G}|\mathcal{G}_0)= \sum_{d}\prod_{a=N_0+1}^{N}C_{ad}\prod_{a=N_0+1}^{N}p_{ad} \textrm{,}
\end{equation}
where $d$ ranges over all sequences of the elementary extensions that are included in the transition from $\mathcal{G}_0$ to $\mathcal{G}$.

In \cite{RideoutSorkin} the sequential growth dynamics of the causal set satisfies the ``discrete general covariance''. This is the condition that any two sequences of the elementary extensions with the same initial d-graph $\mathcal{G}_0$ and final d-graph $\mathcal{G}$ have the same probability. This condition is very strong. Probably such dynamics cannot generate any complicated structures and cannot describe particles. We require here the condition that is apparently much weaker.

\begin{law1}\label{P33} The product of functions $p_{ad}$ in (\ref{eq:sgd4.3}) is the same for any sequences of the elementary extensions with the same initial d-graph $\mathcal{G}_0$ and final d-graph $\mathcal{G}$.
\end{law1}

By definition, put
\begin{equation}
\label{eq:sgd4.4} p(\mathcal{G}|\mathcal{G}_0)=\prod_{a=N_0+1}^{N}p_{ad} \textrm{.}
\end{equation}
We have
\begin{equation}
\label{eq:sgd4.5} P(\mathcal{G}|\mathcal{G}_0)= \left(\sum_{d}\prod_{a=N_0+1}^{N}C_{ad}\right) p(\mathcal{G}|\mathcal{G}_0) \textrm{.}
\end{equation}
In distinction to a \cite{RideoutSorkin} the normalization constants may be different for different sequences of the elementary extensions.

\begin{figure}
	\centering	
		\includegraphics[width=4cm,trim=8cm 18cm 8cm 6cm]{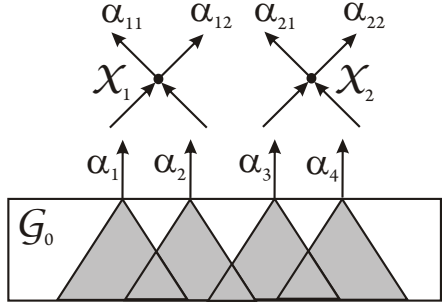}
	\caption{Two internal elementary extensions to the future.}
	\label{fig:fig10}
\end{figure}

\begin{figure}
	\centering	
		\includegraphics[width=4cm,trim=8cm 18cm 8cm 5cm]{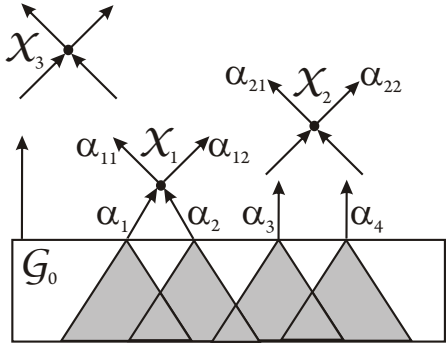}
	\caption{The elementary extensions to $\mathcal{G}_1=\mathcal{G}_0\cup\mathcal{X}_1$.}
	\label{fig:fig11}
\end{figure}

If we consider only the elementary extensions to the future (or to the past) conjecture \ref{P33} is a consequence of conjecture \ref{P32}. Consider a simple example of the sequence of two internal elementary extensions to the future of $\mathcal{G}_0$ (Fig.\ \ref{fig:fig10}). The first elementary extension is the addition of a new x-structure $\mathcal{X}_1$ to the maximal monads $\alpha_1$ and $\alpha_2$. The second elementary extension is the addition of a new x-structure $\mathcal{X}_2$ to the maximal monads $\alpha_3$ and $\alpha_4$. Let $P_1(\mathcal{G}_0)$ and $P_2(\mathcal{G}_0)$ be the probabilities of these elementary extensions to $\mathcal{G}_0$, respectively. We have
\begin{equation}
\label{eq:sgd4.6} P_1(\mathcal{G}_0)=C_0p_1 \textrm{,}
\end{equation}
\begin{equation}
\label{eq:sgd4.7} P_2(\mathcal{G}_0)=C_0p_2 \textrm{.}
\end{equation}
We have the same normalization constant $C_0$ by (\ref{eq:sgd2.10}).
We get the d-graph $\mathcal{G}_1$ by the addition of $\mathcal{X}_1$ to $\mathcal{G}_0$. Consider the addition of $\mathcal{X}_2$ to $\mathcal{G}_1$ (Fig.\ \ref{fig:fig11}). Let $P_2(\mathcal{G}_1)$ be the probability of this elementary extension to $\mathcal{G}_1$. We have
\begin{equation}
\label{eq:sgd4.8} P_2(\mathcal{G}_1)=C_1p_2 \textrm{.}
\end{equation}
We calculate the normalization constant $C_1$ by (\ref{eq:sgd2.10}) as $C_0$. But the sums in (\ref{eq:sgd2.10}) are different for $C_0$ and $C_1$. The sum for $C_0$ includes the additions of new x-structures to the maximal monads $\alpha_1$ and (or) $\alpha_2$. The sum for $C_1$ does not include this elementary extensions and includes the additions of new x-structures $\mathcal{X}_3$ to the maximal monads $\alpha_{11}$ and (or) $\alpha_{12}$ of $\mathcal{X}_1$.

\begin{figure}
	\centering	
		\includegraphics[width=4cm,trim=8cm 18cm 8cm 5cm]{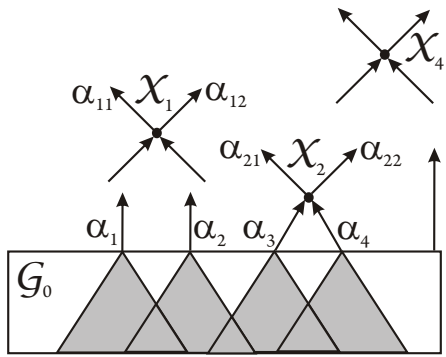}
	\caption{The elementary extensions to $\mathcal{G}_2=\mathcal{G}_0\cup\mathcal{X}_2$.}
	\label{fig:fig12}
\end{figure}

Similarly, we get the d-graph $\mathcal{G}_2$ by the addition of $\mathcal{X}_2$ to $\mathcal{G}_0$. Consider the addition of $\mathcal{X}_1$ to $\mathcal{G}_2$ (Fig.\ \ref{fig:fig12}). Let $P_1(\mathcal{G}_2)$ be the probability of this elementary extension to $\mathcal{G}_2$. We have
\begin{equation}
\label{eq:sgd4.9} P_1(\mathcal{G}_2)=C_2p_1 \textrm{.}
\end{equation}
We calculate the normalization constant $C_2$ by (\ref{eq:sgd2.10}) as $C_0$ and $C_1$. The sums for $C_2$ does not include the additions of new x-structures to the maximal monads $\alpha_1$, and (or) $\alpha_2$, and (or) $\alpha_{11}$, and (or) $\alpha_{12}$. It includes the additions of new x-structures $\mathcal{X}_4$ to the maximal monads $\alpha_{21}$ and (or) $\alpha_{22}$ of $\mathcal{X}_1$.

Let $\mathcal{G}=\mathcal{G}_0\cup\mathcal{X}_1\cup\mathcal{X}_2$. Consider the conditional probability of the transition from $\mathcal{G}_0$ to $\mathcal{G}$. We have
\begin{equation}
\label{eq:sgd4.10} P(\mathcal{G}|\mathcal{G}_0)= C_0C_1p_1p_2+C_0C_2p_2p_1=C_0(C_1+C_2)p_1p_2 \textrm{.}
\end{equation}
This is the particular case of (\ref{eq:sgd4.5}). In the considered example, conjecture \ref{P33} is a consequence of conjecture \ref{P32}. Similarly, we can consider the sequence of internal and external elementary extensions to the future (to the past). The general case follows by induction.

In the general case, $C_0$, $C_1$, and $C_2$ are different. This difference of the normalization constants describes the difference of the sample spaces. The probability describes two sides of the dynamics. On the one hand the probabilities of d-graphs describe the real distribution of the processes in the universe, i.e. the structure of the universe. This is the conjecture \ref{P33}. On the other hand these probabilities describe the properties of measurements. The observer can choose different sample spaces. The sample space describes an information awareness of the observer. The different sequences of measurements correspond to the different sample spaces and consequently the different normalization constants.

\begin{figure}
	\centering	
		\includegraphics[width=4.5cm,trim=8cm 15cm 8cm 6cm]{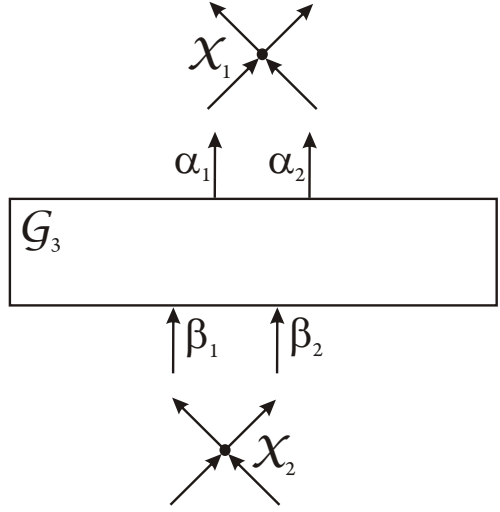}
	\caption{The elementary extensions to the future and to the past.}
	\label{fig:fig13}
\end{figure}

Consider a sequence of two internal elementary extensions of $\mathcal{G}_3$ (Fig.\ \ref{fig:fig13}). In distinction to the previous example these are the elementary extensions to the future and to the past. The first elementary extension to the future is the addition of a new x-structure $\mathcal{X}_1$ to the maximal monads $\alpha_1$ and $\alpha_2$. The second elementary extension to the past is the addition of a new x-structure $\mathcal{X}_2$ to the minimal monads $\beta_1$ and $\beta_2$. Let $P_1(\mathcal{G}_3)$ and $P_2(\mathcal{G}_3)$ be the probabilities of these elementary extensions to $\mathcal{G}_3$, respectively. We have
\begin{equation}
\label{eq:sgd5.1} P_1(\mathcal{G}_3)=C_3p_1(\mathcal{G}_3) \textrm{,}
\end{equation}
\begin{equation}
\label{eq:sgd5.2} P_2(\mathcal{G}_3)=C_3p_2(\mathcal{G}_3) \textrm{.}
\end{equation}

We get the d-graph $\mathcal{G}_4$ by the addition of $\mathcal{X}_1$ to $\mathcal{G}_3$. Consider the addition of $\mathcal{X}_2$ to $\mathcal{G}_4$. Let $P_2(\mathcal{G}_4)$ be the probability of this elementary extension to $\mathcal{G}_4$. We have
\begin{equation}
\label{eq:sgd5.3} P_2(\mathcal{G}_4)=C_4p_2(\mathcal{G}_4) \textrm{.}
\end{equation}
$\mathcal{X}_1$ can be included in the future of $\mathcal{X}_2$ in $\mathcal{G}_4$. In the general case, $p_2(\mathcal{G}_3)\ne p_2(\mathcal{G}_4)$.

Similarly, we get the d-graph $\mathcal{G}_5$ by the addition of $\mathcal{X}_2$ to $\mathcal{G}_3$. Consider the addition of $\mathcal{X}_1$ to $\mathcal{G}_5$. Let $P_1(\mathcal{G}_5)$ be the probability of this elementary extension to $\mathcal{G}_5$. We have
\begin{equation}
\label{eq:sgd5.4} P_1(\mathcal{G}_5)=C_5p_1(\mathcal{G}_5) \textrm{.}
\end{equation}
$\mathcal{X}_2$ can be included in the past of $\mathcal{X}_1$ in $\mathcal{G}_5$. In the general case, $p_1(\mathcal{G}_3)\ne p_1(\mathcal{G}_5)$.

Let $\mathcal{G}_6=\mathcal{G}_3\cup\mathcal{X}_1\cup\mathcal{X}_2$. Consider the conditional probability of the transition from $\mathcal{G}_3$ to $\mathcal{G}_6$. We have
\begin{equation}
\label{eq:sgd5.5} P(\mathcal{G}_6|\mathcal{G}_3)= C_3C_4p_1(\mathcal{G}_3)p_2(\mathcal{G}_4)+C_3C_5p_2(\mathcal{G}_3)p_1(\mathcal{G}_5) \textrm{.}
\end{equation}
By conjecture \ref{P33},
\begin{equation}
\label{eq:sgd5.6} p_1(\mathcal{G}_3)p_2(\mathcal{G}_4)=p_2(\mathcal{G}_3)p_1(\mathcal{G}_5) \textrm{.}
\end{equation}
This is the independent assumption.
\subsection{Dynamical structures \label{DS}}
Consider the properties of the function $p(\mathcal{G}|\mathcal{G}_0)$. The application of conjecture \ref{P32} yields
\begin{equation}
\label{eq:sgd5.7} p(\mathcal{G}|\mathcal{G}_0)=\prod_{a=N_0+1}^{N}p_a \textrm{,}
\end{equation}
where the multipliers $p_a$ can depend only on some structures $\mathcal{S}_i\in\mathcal{G}$ and on the order of elementary extensions.
\begin{equation}
\label{eq:sgd5.8} p_a=p_a(\mathcal{S}_a) \textrm{,}
\end{equation}
Such structure $\mathcal{S}_a$ is called a dynamical structure. The dynamical structure is a finite set of monads with some connections of these monads. The dynamical structure may be a set of chains or a set of paths and so on.

Let's define what we mean by the dependence (\ref{eq:sgd4.5}). $p(\mathcal{G}|\mathcal{G}_0)$ includes the multiplier $p_a(\mathcal{S}_a)$ only if $\mathcal{S}_a\in\mathcal{G}$, $\mathcal{G}_0\in\mathcal{G}$, and $\mathcal{S}_a\notin\mathcal{G}_0$. If $\exists \mathcal{G}$ and $\exists \mathcal{G}_0$ such that $p(\mathcal{G}|\mathcal{G}_0)$ includes the multiplier $p_a(\mathcal{S}_a)$; then $p(\mathcal{G}_1|\mathcal{G}_2)$ includes the multiplier $p_a(\mathcal{S}_a)$ for $\forall \mathcal{G}_1$ and $\forall \mathcal{G}_2$ if $\mathcal{S}_a\in\mathcal{G}_1$, $\mathcal{G}_2\in\mathcal{G}_1$, and $\mathcal{S}_a\notin\mathcal{G}_2$. The functions $p_a(\mathcal{S}_a)$ may be different for different $\mathcal{S}_a$.

The dynamical causality and the discrete general covariance restrict the possible form of $\mathcal{S}_a$.

\begin{law2}\label{T1} Let $\mathcal{S}$ be a dynamical structure and $\mathcal{S}\ne\emptyset$; then there is an unique pair of x-structures $\mathcal{X}_a$ and $\mathcal{X}_b$ such that $\exists\gamma_i((\gamma_i\in\mathcal{X}_a)\wedge(\gamma_i\in S))$, $\exists\gamma_j((\gamma_j\in\mathcal{X}_b)\wedge(\gamma_j\in S))$, $\mathcal{A}(\mathcal{X}_a,\ \mathcal{X}_b)\ne\emptyset$, and $\mathcal{S}\in(\mathcal{A}(\mathcal{X}_a,\ \mathcal{X}_b)\cup\mathcal{X}_a \cup\mathcal{X}_b)$.
\end{law2}

Proof: $\mathcal{S}$ is a finite set of monads $\{\gamma_l(S)| \gamma_l(S)\in \mathcal{S}\}$. Each monad is included in the x-structure. Consider a set of x-structures $\{\mathcal{X}_s(\mathcal{S})| \gamma_l(\mathcal{S})\in\mathcal{X}_s(\mathcal{S})\}$. This set is a d-graph $\mathcal{G}_s$. We have $\mathcal{S}\in\mathcal{G}_s$. Hence the probability $P(\mathcal{G}_s)$ includes the factor $p(\mathcal{S})$.

Let $\gamma_1$ be a maximal monad of $\mathcal{S}$. Consider an x-structure $\mathcal{X}_1$ such that $\gamma_1\in\mathcal{X}_1$. Let $\mathcal{G}_{s1}=\mathcal{G}_s\setminus\mathcal{X}_1$. We have $\mathcal{S}\notin\mathcal{G}_{s1}$. Hence the probability $P(\mathcal{G}_{s1})$ does not include the factor $p(\mathcal{S})$.

The probability $P(\mathcal{G}_s|\mathcal{G}_{s1})$ is the probability of the adding of $\mathcal{X}_1$ to $\mathcal{G}_{s1}$. This is the elementary extension to the future. We have
\begin{equation}
\label{eq:sgd5.9} P(\mathcal{G}_s|\mathcal{G}_{s1})= P(\mathcal{G}_s)/P(\mathcal{G}_{s1})\propto p(\mathcal{S}) \textrm{,}
\end{equation}
Consequently $S\in (\mathcal{P}(\mathcal{X}_1) \cup\mathcal{X}_1)$.

Similarly let $\gamma_2$ be a minimal monad of $\mathcal{S}$. Consider an x-structure $\mathcal{X}_2$ such that $\gamma_2\in\mathcal{X}_2$. Let $\mathcal{G}_{s2}=\mathcal{G}_s\setminus\mathcal{X}_2$. We have $S\notin\mathcal{G}_{s2}$. Hence the probability $P(\mathcal{G}_{s2})$ does not include the factor $p(\mathcal{S})$.

The probability $P(\mathcal{G}_s|\mathcal{G}_{s2})$ is the probability of the adding of $\mathcal{X}_2$ to $\mathcal{G}_{s2}$. This is the elementary extension to the past. We have
\begin{equation}
\label{eq:sgd5.10} P(\mathcal{G}_s|\mathcal{G}_{s2})= P(\mathcal{G}_s)/P(\mathcal{G}_{s2})\propto p(\mathcal{S}) \textrm{,}
\end{equation}
Consequently $\mathcal{S}\in\mathcal{F}(\mathcal{X}_2) \cup\mathcal{X}_2$. 

By definition, put $\mathcal{X}_2=\mathcal{X}_a$, $\mathcal{X}_1=\mathcal{X}_b$, $\gamma_2=\gamma_i$, and $\gamma_1=\gamma_j$. We get $\mathcal{S}\in ((\mathcal{P}(\mathcal{X}_b) \cup\mathcal{X}_b) \cap (\mathcal{F}(\mathcal{X}_a) \cup\mathcal{X}_a))$. We have $(\mathcal{P}(\mathcal{X}_b) \cap \mathcal{F}(\mathcal{X}_a))=\mathcal{A}(\mathcal{X}_a,\ \mathcal{X}_b)$. If $\mathcal{A}(\mathcal{X}_a,\ \mathcal{X}_b)=\emptyset$ $((\mathcal{P}(\mathcal{X}_b) \cup\mathcal{X}_b) \cap (\mathcal{F}(\mathcal{X}_a) \cup\mathcal{X}_a))=\emptyset$, and $\mathcal{S}=\emptyset$. This is a contradiction. If $\mathcal{A}(\mathcal{X}_a,\ \mathcal{X}_b)\ne\emptyset$ $\mathcal{X}_a\in\mathcal{P}(\mathcal{X}_b)$, and $\mathcal{X}_b\in\mathcal{F}(\mathcal{X}_a)$. We get $\mathcal{S}\in ((\mathcal{P}(\mathcal{X}_b) \cup\mathcal{X}_b) \cap (\mathcal{F}(\mathcal{X}_a) \cup\mathcal{X}_a))= (\mathcal{A}(\mathcal{X}_a,\ \mathcal{X}_b)\cup\mathcal{X}_a \cup\mathcal{X}_b)$. $\Box$

Any factor $p(\mathcal{S})$ is a function of the pair of x-structures. If the considered d-graph has $N$ x-structures we can represent all these factors as a square matrix of size $N$. The matrix representation of this dynamics will be considered in following papers.
\section{EMERGENT STRUCTURES \label{ES}}
\subsection{Binary alternatives and loops \label{BA}}
Our goal in this section will be to propose a particular simple example of the consequence growth dynamics. The dynamics is complete if there is a law of calculation of probabilities of elementary extensions for any d-graph. This law is an equation of motion. The particular example of the dynamics means the particular example of $\mathcal{S}$ and $p(\mathcal{S})$. To find them we need some additional conjectures.

In the consequence growth dynamics a probability is a primordial entity. One of the main idea of this approach is the discreteness on a microscopic scale. We can apply this idea to a probability. ``It is certainly possible to decide any large alternative step by step in binary alternatives.'' \cite[p.\ 222]{vonW1975}. It is attractive to take the binary alternative as a primordial entity of a discrete pregeometry \cite[\S\ 44.5]{Gr}.

The binary alternative is a decision with two equiprobable outcomes. This is a fork in an progress of a process. In a d-graph such fork is any x-structure. We can consider a path in the d-graph as an progress of some process. In each x-structure this process is confronted by a binary decision: forward to the left or forward to the right. This choice has the probabilities $(1/2)\times(1/2)$ by a symmetry. Hence the path has the probability $2^{-m}$ if it includes $m$ chronons. The d-graph includes the whole of x-structure, the both outcomes. The sum of the probabilities of all paths from any x-structure is equal to 1. The nontrivial structure is an intersection of two paths.

\begin{figure}
	\centering	
		\includegraphics[width=4cm,trim=8cm 17cm 8cm 5cm]{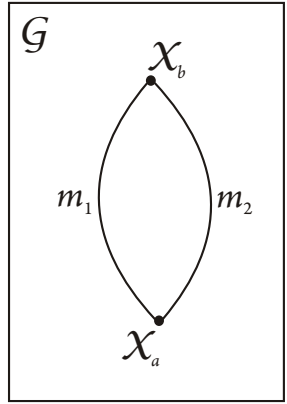}
	\caption{The simple loop.}
	\label{fig:fig14}
\end{figure}

Consider two x-structures $\mathcal{X}_a$ and $\mathcal{X}_b$. Suppose there are only two paths such that the start monads of the both paths belong to $\mathcal{X}_a$, the final monads of the both paths belong to $\mathcal{X}_b$, and these paths have not common monads (Fig.\ \ref{fig:fig14}). This structure is called a simple loop $\mathcal{L}$. By definition, assume $\mathcal{L}$ is a dynamical structure and $p(\mathcal{L})=2^{-m_1}2^{-m_2}$ if these paths have $m_1$ and $m_2$ chronons, respectively. This assumption has clear physical meaning. The probability of the interaction of two processes is high if their common past was recently. This probability decreases exponentially if their common past was long ago.

\begin{figure}
	\centering	
		\includegraphics[width=4cm,trim=8cm 14cm 8cm 6cm]{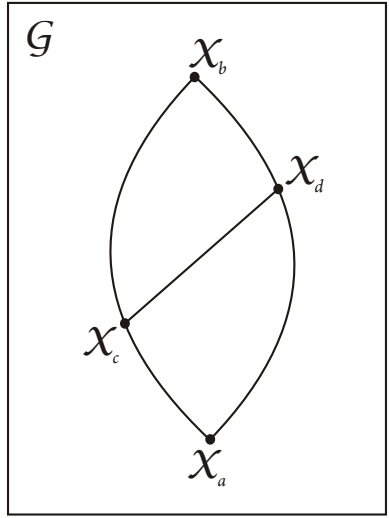}
	\caption{Three paths between the x-structures.}
	\label{fig:fig15}
\end{figure}

Consider the modified structure. Suppose there are only three paths such that the start monads of these paths belong to $\mathcal{X}_a$, the final monads of these paths belong to $\mathcal{X}_b$ (Fig.\ \ref{fig:fig15}). The first and second path have not common monads as in the previous example. The segments of first and third path coincide between $\mathcal{X}_a$ and $\mathcal{X}_c$. The segments of second and third path coincide between $\mathcal{X}_d$ and $\mathcal{X}_b$. This structure consists of two simple loops. The first simple loop is between $\mathcal{X}_a$ and $\mathcal{X}_d$. The second simple loop is between $\mathcal{X}_c$ and $\mathcal{X}_b$. These simple loops are dynamical structures. The intersection of the first and second path in $\mathcal{X}_b$ is already included in the second simple loop. Hence assume that the pair of the x-structures $\mathcal{X}_a$ and $\mathcal{X}_b$ does not generate any dynamical structure.

Consider the general case. Two paths are called connected paths if there is a pair of causally connected monads such that the one monad is included in the first path and the second monad is included in the second path. The possible structures between two x-structures are described by the following theorem.

\begin{figure}
	\centering	
		\includegraphics[width=4cm,trim=8cm 16cm 8cm 4cm]{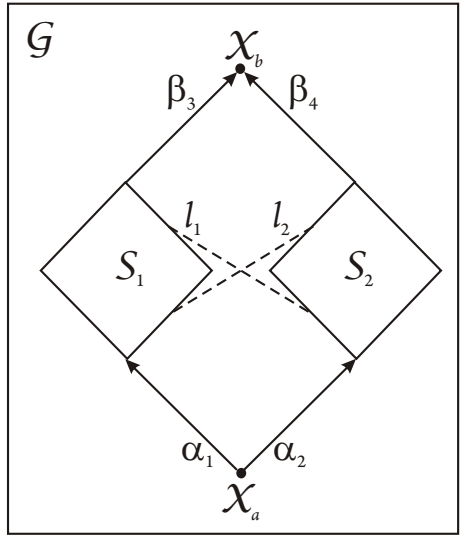}
	\caption{The loop.}
	\label{fig:fig16}
\end{figure}

\begin{law2}\label{T2} If there are more than one path between two x-structures $\mathcal{X}_a$ and $\mathcal{X}_b$, then only two case are possible:
\begin{itemize}
\item all paths between $\mathcal{X}_a$ and $\mathcal{X}_b$ are connected;
\item there are two groups of paths between $\mathcal{X}_a$ and $\mathcal{X}_b$; all paths in each group are connected, and each path of one group is not connected with each path of second group.
\end{itemize}
The structure in the second case is called a loop $\mathcal{L}_{ab}$ (Fig.\ \ref{fig:fig16}). $\mathcal{X}_a$ is called the start x-structure. $\mathcal{X}_b$ is called the finish x-structure.
\end{law2}

Proof: $\mathcal{X}_a$ includes two monads of creation $\alpha_1$ and $\alpha_2$. If all paths start in one of these monads we have the first case. Similarly $\mathcal{X}_b$ includes two monads of destruction $\beta_3$ and $\beta_4$. If all paths finish in one of these monads we have the first case. Consider a group $\mathcal{S}_1$ of paths that have the same start monad $\alpha_1$ and the same finish monad $\beta_3$. This is a group of connected paths. Similarly consider a group $\mathcal{S}_2$ of paths that have the same start monad $\alpha_2$ and the same finish monad $\beta_4$. This is also a group of connected paths. Suppose there are two monads $\gamma_1$ and $\gamma_2$ such that $\gamma_1\in\mathcal{S}_1$, $\gamma_2\in\mathcal{S}_2$, and $\gamma_1\prec\gamma_2$. In this case there is a path from $\alpha_1$ to $\beta_4$. This path includes three segments. The first segment is the path from $\alpha_1$ to $\gamma_1$. The second segment is the path from $\gamma_1$ to $\gamma_2$ (the dashed line $l_2$ in Fig.\ \ref{fig:fig16}). The third segment is the path from $\gamma_2$ to $\beta_4$. The common monad $\alpha_1$ of the paths of $\mathcal{S}_1$ is causally connected with the common monad $\beta_4$ of the paths of $\mathcal{S}_2$. Consequently all paths are connected. Similarly suppose there are two monads $\gamma_3$ and $\gamma_4$ such that $\gamma_3\in\mathcal{S}_1$, $\gamma_4\in\mathcal{S}_2$, and $\gamma_4\prec\gamma_3$. In this case there is a path from $\alpha_2$ to $\beta_3$ (the dashed line $l_1$ in Fig.\ \ref{fig:fig16}) and all paths are connected. If there are not the paths from $\alpha_1$ to $\beta_4$ and from $\alpha_2$ to $\beta_3$ there is a loop between $\mathcal{X}_a$ and $\mathcal{X}_b$. This is all possible cases. $\Box$

Assume a following conjecture for simplicity.

\begin{law1}\label{P41} The group of connected paths is not a dynamical structure. The loop is a dynamical structure. The loop $\mathcal{L}$ generate the factor
\begin{equation}
\label{eq:es1.1} p(\mathcal{L})=2^{-m} \textrm{,}
\end{equation}
where $m$ is the number of the chronons in $\mathcal{L}$.
\end{law1}

Consider the dependence of the probability of the elementary extension on the loops. This is the conditional probability of the transition from $\mathcal{G}_N$ to $\mathcal{G}_{N+1}$, where $N$ is the number of x-structures in $\mathcal{G}_N$. We have
\begin{equation}
\label{eq:es1.2} P(\mathcal{G}_{N+1}|\mathcal{G}_N)= C p(\mathcal{S}) \sum_L p(\mathcal{L}) \textrm{,}
\end{equation}
where $C$ is the normalization constant, $p(\mathcal{S})$ describes the possible dependence on the other structures $S$ that is not loops. There is the summation of all new loops that appear by this elementary extension. The external elementary extension does not generate loops. The internal elementary extension can generate several loops. New x-structure is a start x-structure of new loops if this is the internal elementary extension to the past. New x-structure is a finish x-structure of new loops if this is the internal elementary extension to the future. These loops satisfy the following theorem.

\begin{law2}\label{T3} Consider the internal elementary extension to the future (to the past). This is the addition of an x-structure $\mathcal{X}_b$. If this elementary extension generates several loops with the set $\{\mathcal{X}_a\}$ of start (finish) x-structures, then the set of monads of one kind (the creation or the annihilation) that belong to $\{\mathcal{X}_a\}$ is a antichain.
\end{law2}

Proof: The proof is by reductio ad absurdum. Consider the internal elementary extension to the future, and the set of monads $\{\alpha_i\}$ that belong to $\{\mathcal{X}_a\}$. The proof is the same for the internal elementary extension to the past or for the set of monads of the annihilation. Assume the converse, then there are two monads $\alpha_1$ and $\alpha_2$, $\alpha_1\in\mathcal{X}_1\in\{\mathcal{X}_a\}$, $\alpha_2\in\mathcal{X}_2\in\{\mathcal{X}_a\}$, and $\alpha_1\prec\alpha_2$. If $\alpha_1\prec\alpha_2$, then $\exists\alpha_3$, $\alpha_3\in\mathcal{X}_2\in\{\mathcal{X}_a\}$, and $\alpha_1\prec\alpha_3$. We have a path from $\alpha_1$ to $\alpha_2$ and a path from $\alpha_1$ to $\alpha_3$. Under the conditions of the theorem $L_{2b}$ is a loop, then there are a path from $\alpha_2$ to $\beta_{b1}$ and there are a path from $\alpha_3$ to $\beta_{b2}$, where $\beta_{b1}\in\mathcal{X}_b$ and $\beta_{b2}\in\mathcal{X}_b$. We have two paths. Each path begins in the same monad $\alpha_1$ but comes to an end in the different monads $\beta_{b1}$ and $\beta_{b2}$. Consequently all paths from $\mathcal{X}_1$ to $\mathcal{X}_b$ are connected as in the proof of theorem\ \ref{T2}. The x-structures $\mathcal{X}_1$ and $\mathcal{X}_b$ do not generate a loop. This contradiction proves the theorem. $\Box$

The physical meaning of a loop is one cycle of a cyclic process. The future dynamics of processes depends only on last cycles. This is the memory depth. The dynamics does not depend on the structure of loops, and depends only on the number of monads. This property corresponds to the hierarchical structure of the matter. The properties of high levels do not depend on the detailed structure of low levels. The probability decreases exponentially in dependence on the size of the loop. In the real world the dependence of the probability on the loop may be more complicated. But the main properties remain changeless.
\subsection{Model 1 \label{M1}}
Consider a first simple example. Assume the following conditions:
\begin{itemize}
\item the probabilities of external elementary extensions are equal to 0;
\item there is condition (\ref{eq:sgd2.90});
\item the probabilities depend only on the loops.
\end{itemize}
The first condition means that the considered system is a closed world. The considered d-graph has the constant width.

\begin{figure}
	\centering	
		\includegraphics[width=4.5cm,trim=8cm 19cm 8cm 6cm]{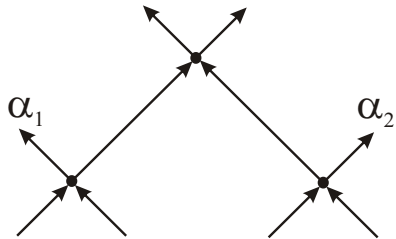}
	\caption{The pair of maximal monads with disjoint past sets.}
	\label{fig:fig17}
\end{figure}

Let $\mathcal{G}_0$ is a initial d-graph. $\mathcal{G}_0$ can include a pair of maximal monads with disjoint past sets or a pair of minimal monads with disjoint future sets. For instance, in the Fig.\ \ref{fig:fig17} such pair of maximal monads is $\alpha_1$ and $\alpha_2$. An internal elementary extension of such pair does not generate loops. Hence the probability of this elementary extension has the maximal value. After this elementary extension, new maximal monads belong to one x-structure and their past sets coincide. Obviously, internal elementary extensions can not generate pairs of maximal monads with disjoint past sets. $\mathcal{G}_0$ can include only finite set of such pairs. The d-graph will not include such pairs after the finite sequence of internal elementary extensions. Similarly, the d-graph will not include the pair of minimal monads with disjoint future sets.

\begin{figure}
	\centering	
		\includegraphics[width=4cm,trim=8cm 18cm 8cm 3cm]{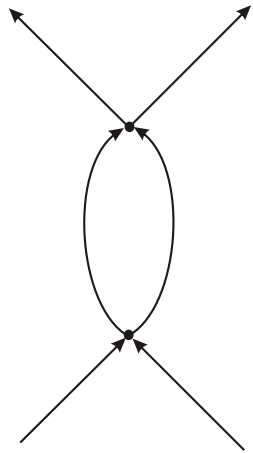}
	\caption{The minimal loop.}
	\label{fig:fig18}
\end{figure}

\begin{figure}
	\centering	
		\includegraphics[width=3.8cm,trim=8cm 11cm 8cm 6.3cm]{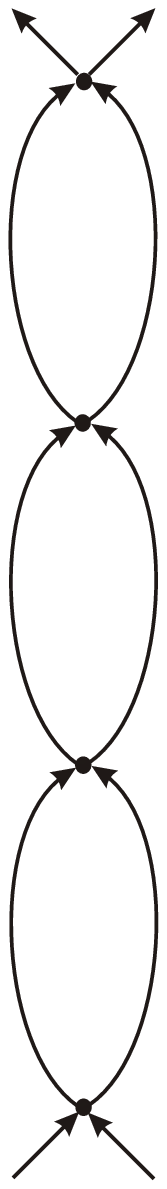}
	\caption{The sequence of minimal loops.}
	\label{fig:fig19}
\end{figure}

\begin{figure}
	\centering	
		\includegraphics[width=3.8cm,trim=8cm 3cm 8cm 4cm]{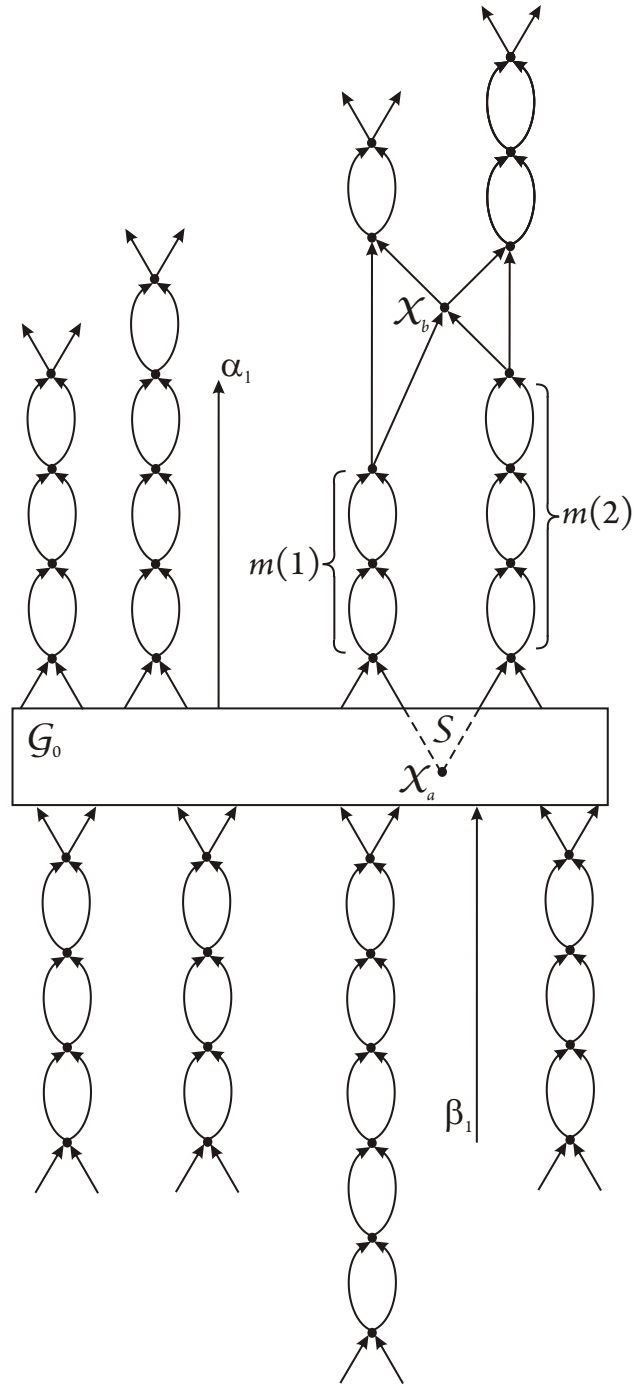}
	\caption{The evolution of the closed world.}
	\label{fig:fig20}
\end{figure}

Consider the d-graph without the pairs of maximal monads with disjoint past sets or the pairs of minimal monads with disjoint future sets. The probability of the elementary extensions has the maximal value if this extension generate one minimal loop. This loop includes two chronons (Fig.\ \ref{fig:fig18}). By $\mathcal{\tilde L}$ denote a minimal loop. $p(\mathcal{\tilde L})=1/4$. The sequence of the internal elementary extensions can generate the sequence of $\mathcal{\tilde L}$ (Fig.\ \ref{fig:fig19}). The width of this d-graph is equal to 2.
The probabilities of other loops decrease exponentially depending on the numbers of the chronons in these loops. We get only minimal loops as a result of the consequential growth dynamics for any $\mathcal{G}_0$ after some transient process (Fig.\ \ref{fig:fig20}). For instance, consider the elementary extension that generates $\mathcal{L}_{ab}$. We have
\begin{equation}
\label{eq:es1.3} P(\mathcal{L}_{ab})= C 2^{-m(\mathcal{S})-m(1)-m(2)-2} \textrm{,}
\end{equation}
where $C$ is the normalization constant, $m(\mathcal{S})$ is the number of chronons in the beginning structure $\mathcal{S}$ of $\mathcal{L}_{ab}$, and $m(1)$ and $m(2)$ is the numbers of chronons in the left and right sequences of $\mathcal{\tilde L}$, respectively. In the course of sequential growth the lengths of sequences of $\mathcal{\tilde L}$ increase and the probabilities of such loops decrease exponentially. If the width of $\mathcal{G}_0$ is odd, there are the redundant maximal and minimal monads ($\alpha_1$ and $\beta_1$).

The considered universe consists of particles of two kinds. The first kind is the redundant maximal and minimal monads. These are the noninteracting chronons. These particles have not internal time. In the continuous limit these particles must correspond to massless particles. The second kind is the sequences of $\mathcal{\tilde L}$. The minimal loop is a minimal cycle. In the continuous limit these particles must correspond to heavy particles. Let's call they minimons.

Any closed world disintegrates to noninteracting simplest elementary particles. This is the heat death of the universe.
\subsection{Model 2 \label{M2}}
Consider a second simple example. The second and third conditions are the same. But the probabilities of external elementary extensions are not equal to 0. This means the considered system is open. We have the condition (\ref{eq:sgd2.91}).

Using \ref{eq:sgd2.7} - \ref{eq:sgd2.91}, we get
\begin{equation}
\label{eq:es1.4} P_{ext,f}=P_{ext,p}=\frac{1}{2(1+K)} \textrm{.}
\end{equation}
The external elementary extension does not generate loops. Using \ref{eq:sgd2.3} and \ref{eq:sgd2.5}, we get
\begin{equation}
\label{eq:es1.5} P_{i(ext,f)}=P_{i(ext,p)}=\frac{1}{2n(1+K)} \textrm{,}
\end{equation}
where $n$ is the width of the d-graph.

Consider the case $K\ll 1$. Almost all elementary extensions are external. These extensions generate the pairs of maximal monads with disjoint past sets and the pairs of minimal monads with disjoint future sets. Hence most internal elementary extensions do not generate the loops. These are rare elementary extensions. The probability of the appearance of $\mathcal{\tilde L}$ has the factor $1/4$. The probability of the appearance of the big loops is exponentially small. This d-graph has not the regular structure. We can call it the disordered phase or the chaos. The chaos lacks long-range structure, but still has causal order. This case describes the very intensive influence of the environment. This is the superdense phase such that the existence of particles is impossible.

Our goal is to describe the particles in the vacuum. Below we shall consider only the case $K\gg 1$.

\begin{figure}
	\centering	
		\includegraphics[width=3.8cm,trim=8cm 1.5cm 8cm 4cm]{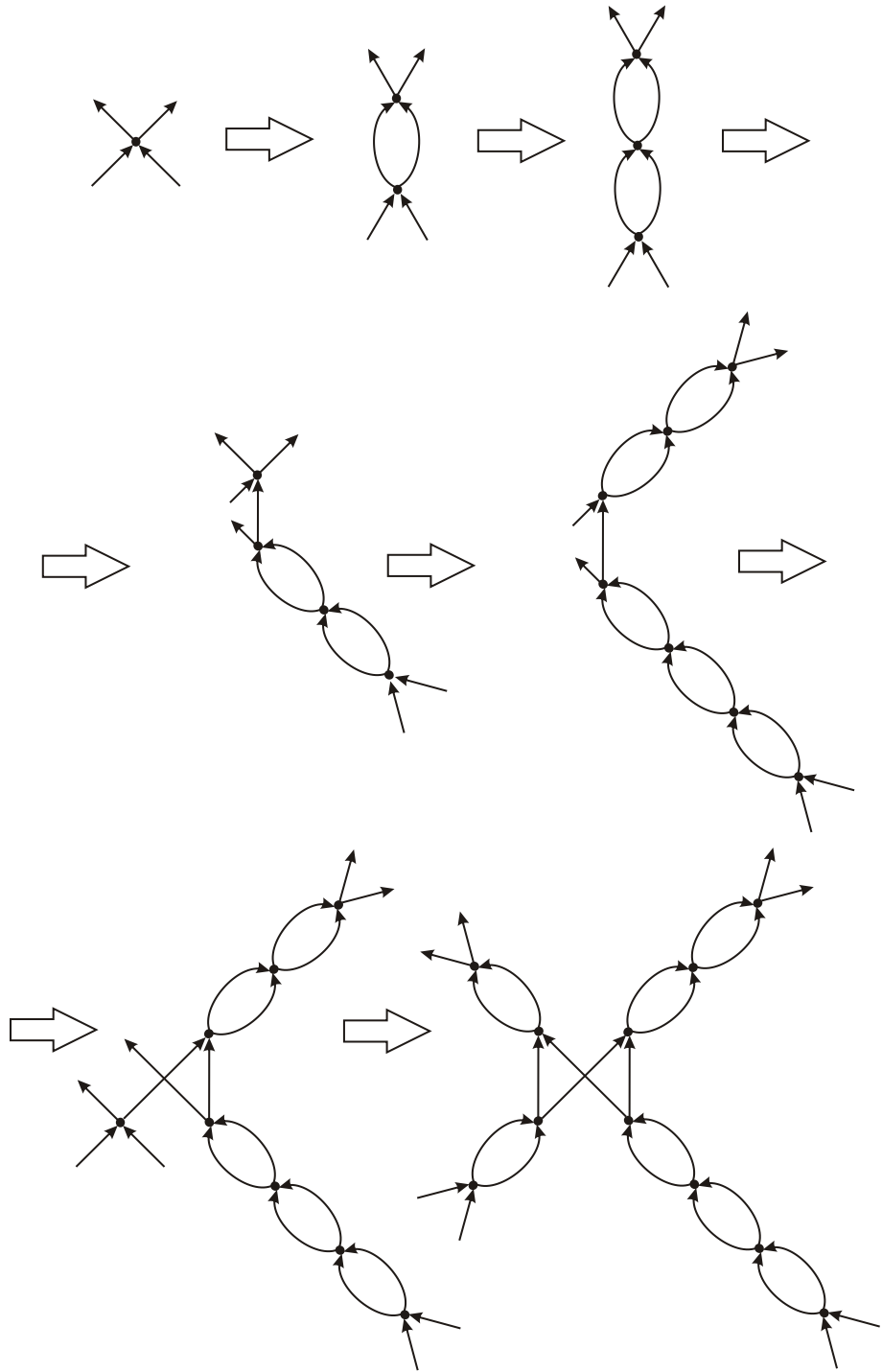}
  \caption{The growth of d-graph from the empty set.}
	\label{fig:fig21}
\end{figure}

Consider the growth of d-graph from the empty set (Fig.\ \ref{fig:fig21}). The first elementary extension generates the x-structure with probability 1. Then we have the growth of the sequence of $\mathcal{\tilde L}$ with high probability up to the first external elementary extension. The average number of $\mathcal{\tilde L}$ in this sequence is the number of failures up to the first success in the sequence of Bernoulli tests. It equal to the ratio $K$ of probabilities of the internal and external elementary extensions. After the first external elementary extension the ends of the sequence continue the growth. This structure is repeated if the next external elementary extension appears in the end of the sequence of $\mathcal{\tilde L}$. This is not in the figure. If the next external elementary extension appears close by the first we have two new growing sequences of $\mathcal{\tilde L}$. This structure describes the collision of two minimons. The further growth describes the increasing number of the collided minimons. The physical meaning of $K$ is the average time of free path. The time means the number of cycles. The repeated collision is the appearance of the big loop. It probability is exponentially small.

This is the realistic model of the infinite volume of the gas of the minimons. But it does not generate the hierarchy of structures. We need the attraction of the minimons.
\subsection{Model 3 \label{M3}}
The attraction means the increasing of the probability of some new connections. The fundamental connection in this approach is a causal connection. Assume the following conjecture.

\begin{law1}\label{P42} The pair of causally connected external monads is a dynamical structure. This dynamical structure generates the factor
\begin{equation}
\label{eq:es4.1} p(\mathcal{S})=2^q \textrm{,}
\end{equation}
where $q\gg 1$.
\end{law1}

The pair of causally connected external monads consists of minimal and maximal monad.

Consider the modified model. We take into account the influence of causally connected external monads. Other conditions are the same as in the previous case:
\begin{itemize}
\item there are conditions (\ref{eq:sgd2.90}) and (\ref{eq:sgd2.91});
\item $K\gg 1$;
\item the probabilities depend only on the loops and the causally connected external monads.
\end{itemize}
Any internal elementary extension can generate new causal connection between the external monads, but it can not generate the causally disconnected pairs of external monads. If the probabilities of the external elementary extensions are equal to 0 as in the first model all pairs of external monads will be causally connected after some transient process. Then the causal connections of the external monads will be the same for all internal elementary extensions. The evolution of the d-graph will be the same as in the first model.

\begin{figure}
	\centering	
		\includegraphics[width=4cm,trim=8cm 16cm 8cm 4cm]{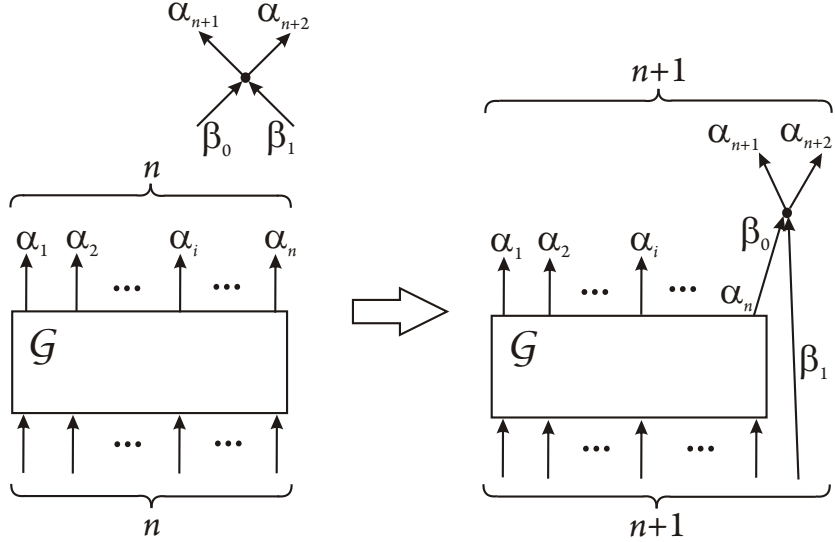}
	\caption{The external elementary extension to the future generates the additional pairs of causally connected and disconnected external monads.}
	\label{fig:fig22}
\end{figure}

Any external elementary extension generates two additional pairs of causally connected external monads ($\beta_1\prec\alpha_{n+1}$ and $\beta_1\prec\alpha_{n+2}$ in Fig.\ \ref{fig:fig22}). These monads are included in the new x-structure. The factor $2^{2q}$ of these pairs can be included in the normalization constants (\ref{eq:sgd2.11}) and (\ref{eq:sgd2.13}). Any external elementary extension generates $n-1$ additional pairs of causally disconnected external monads ($\beta_1\nprec\alpha_i$ in Fig.\ \ref{fig:fig22}, where $i$ ranges from 1 to $n-1$). If the initial external monad ($\alpha_n$ in Fig.\ \ref{fig:fig22}) is included in $k$ pairs of causally connected external monads the external elementary extension generates $k$ additional pairs of causally connected external monads. There are $2k$ new pairs with the monads $\alpha_{n+1}$ and $\alpha_{n+2}$ instead of $k$ pairs with the monad $\alpha_n$. In distinction to the previous model the external elementary extensions have different probabilities.

Particles can not appear from the vacuum. Any particles appear from other particles. In this approach particles are parts of a d-graph with some symmetry. Hence such regular structures must appear by the sequential growth dynamics from other regular structures and can not appear from a stochastic d-graph. A stable particle is a repetitive (cyclic) regular structure. This structure must repeat itself with probability that is close to 1. Such sequential growth dynamics is almost deterministic process. Two examples of simplest cyclic regular structures are a chronon and a minimon. Consider a more complicated example.

\begin{figure}
	\centering	
		\includegraphics[width=12cm,trim=1cm 0cm 1cm 0cm]{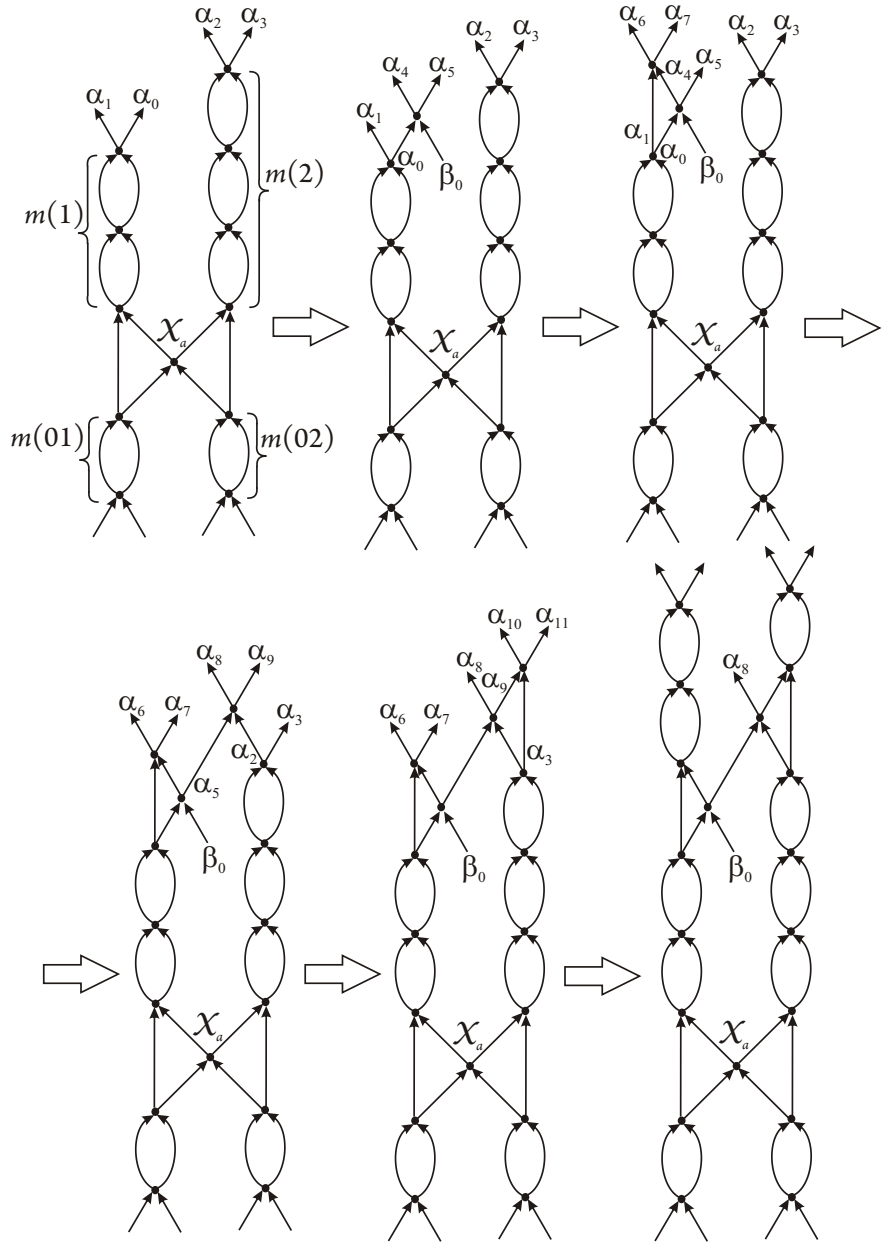}
	\caption{The appearance of the big loop.}
	\label{fig:fig23}
\end{figure}

The initial d-graph consists of two sequences of $\mathcal{\tilde L}$ (Fig.\ \ref{fig:fig23}). Let the numbers $m(01)$ and $m(02)$ of chronons in these sequences be very big: $m(01)\gg K$ and $m(02)\gg K$. In the figure there are only one $\mathcal{\tilde L}$ in each sequence for simplicity. We may take away the consideration of elementary extensions to the past by this assumption. These sequences interact in the x-structure $\mathcal{X}_a$.

The first stage of the sequential growth is the growth of two sequences of $\mathcal{\tilde L}$ to the future. The average numbers $m(1)$ and $m(2)$ of chronons in these sequences have the order of magnitude $K$. Then we have the first external elementary extension. The new minimal monad $\beta_0$ is causally disconnected with maximal monads $\alpha_1$, $\alpha_2$, and $\alpha_3$. Then the internal elementary extension generates the loop with three chronons and one additional pair of causally connected external monads. By assumption, $q\gg 1$, and the probability of this extension is close to 1. The next internal extension can generate one additional pair of causally connected external monads only if it generates a big loop. This is the addition of a new x-structure to $\alpha_5$ and $\alpha_2$ (or $\alpha_3$). The addition of a new x-structure to $\alpha_6$ or $\alpha_7$ is less by the factor $1/4$. We have
\begin{equation}
\label{eq:es4.2} P(\mathcal{L})=C2^{q-5-m(1)-m(2)} \textrm{,}
\end{equation}
where $P(\mathcal{L})$ is the probability of this elementary extension. If $ q\gg5+m(1)+m(2)$ this probability is close to 1. The next internal elementary extension generates the second loop with three chronons and one additional pair of causally connected external monads. This probability is also close to 1. In the result all pairs of external monads are causally connected. Then we have the growth of the two sequences of $\mathcal{\tilde L}$ to the future. This growth goes up to the next external elementary extension.

This big loop is a cycle of the next level that consists of the cycles $\mathcal{\tilde L}$ of the previous level. The considered example shows the existence of the hierarchy of cyclic structures. The next aim is an identification of such structures and different particles. This is the subject for a future work. The computer simulation will be very useful. Each external elementary extension increases a width of a d-graph. This is an expansion of an investigated area of the universe. Consequently a structure of a whole d-graph can not repeat itself. We must pick out cyclic substructures.

In the considered model we have two free parameters: $q$ and $K$. If $q=q(K)$ there is only one free parameter $K$. This parameter describes the influence of an environment and determines the properties of structures. This is Machs's principle in the strongest possible form.

Let's write the equation for the probability of elementary extension.
\begin{equation}
\label{eq:es4.3} P=C\exp(\ln2(kq-\sum_{i}m_i)) \textrm{,}
\end{equation}
where $k$ is the number of additional pairs of causally connected external monads and $m_i$ is the number of chronons in the additional loops that are generated by this elementary extension. We have the summation of these additional loops.
	
This is an equation of motion for the considered model. In more realistic model the equation of motion may be different. But in any case, the dynamical law is not differential but algebraic.
\section{ALGEBRAIC REPRESENTATION \label{AR}}
Here we give an algebraic technique. The aim is to find a mathematical description that is connected with quantum theory in the sense of correspondence principle.
\subsection{State vectors of loops \label{AR1}}
We denote the creation and destruction Bose operators by $\hat a^{\dag}$ and $\hat a$, respectively, satisfying the usual bosonic relations. Consider a state vector $|\mathcal{S}\rangle $ for any set $\mathcal{S}$ of chronons. By definition, put
\begin{equation}
\label{eq:ar1.1} |\mathcal{S}\rangle=\langle 0 | (\hat a)^m(\hat a^{\dag})^m | 0\rangle^{-1/2} (\hat a^{\dag})^m |0\rangle=(m!)^{-1/2} (\hat a^{\dag})^m| 0\rangle\textrm{,}
\end{equation}
where $m$ is the number of chronons in $\mathcal{S}$. We have
\begin{equation}
\label{eq:ar1.2} m=\langle \mathcal{S} | \hat N | \mathcal{S}\rangle \textrm{,}
\end{equation}
where
\begin{equation}
\label{eq:ar1.3} \hat N=\hat a^{\dag}\hat a
\end{equation}
is a particle number operator for chronons.

Consider a d-graph $\mathcal{G}$ and two x-structures $\mathcal{X}_i\equiv \{\alpha_{i1}, \alpha_{i2}, \beta_{i1}, \beta_{i2}\}\in\mathcal{G}$ and $\mathcal{X}_j\equiv \{\alpha_{j1}, \alpha_{j2}, \beta_{j1}, \beta_{j2}\}\in\mathcal{G}$. By $\mathcal{A}_{ij}$ denote an Alexandrov set of chronons. $\mathcal{A}_{ij}=\{(\alpha_a\beta_b)\in\mathcal{G}\mid\beta_{i1}\prec\alpha_a\prec \beta_b \prec\alpha_{j1}\}$. Obviously, $\mathcal{A}_{ij}=\{(\alpha_a\beta_b) \in\mathcal{G}\mid\beta_{i1}\prec\alpha_a\prec \beta_b \prec\alpha_{j2}\}=\{(\alpha_a\beta_b) \in\mathcal{G}\mid\beta_{i2}\prec\alpha_a\prec \beta_b \prec\alpha_{j1}\}=\{(\alpha_a\beta_b) \in\mathcal{G}\mid\beta_{i2}\prec\alpha_a\prec \beta_b \prec\alpha_{j2}\}$. Consider a state vector $| \mathcal{A}_{ij}\rangle$ of $\mathcal{A}_{ij}$. We have
\begin{equation}
\label{eq:ar1.4} m=\langle \mathcal{A}_{ij} | \hat N | \mathcal{A}_{ij}\rangle \textrm{,}
\end{equation}
where $m$ is the number of chronons in $\mathcal{A}_{ij}$. If $\mathcal{A}_{ij}$ is a loop, $m$ is the number of chronons in the loop. In subsection\ \ref{BA} we define a loop as a set of monads. Now we consider a loop as a set of chronons. We use the same notations in both case for simplicity.

Let's determine when $\mathcal{A}_{ij}$ is a loop. By definition, put
\[
\mathcal{A}_{ij (11)}=\{(\alpha_a\beta_b)\in\mathcal{G}\mid\alpha_{i1}\preccurlyeq\alpha_a\prec \beta_b \preccurlyeq\beta_{j1}\}\textrm{,}
\]
\[
\mathcal{A}_{ij (12)}=\{(\alpha_a\beta_b)\in\mathcal{G}\mid\alpha_{i1}\preccurlyeq\alpha_a\prec \beta_b \preccurlyeq\beta_{j2}\}\textrm{,}
\]
\[
\mathcal{A}_{ij (21)}=\{(\alpha_a\beta_b)\in\mathcal{G}\mid\alpha_{i2}\preccurlyeq\alpha_a\prec \beta_b \preccurlyeq\beta_{j1}\}\textrm{,}
\]
\[
\mathcal{A}_{ij (22)}=\{(\alpha_a\beta_b)\in\mathcal{G}\mid\alpha_{i2}\preccurlyeq\alpha_a\prec \beta_b \preccurlyeq\beta_{j2}\}\textrm{.}
\]
$\mathcal{A}_{ij}$ is a loop if
\[
[(\mathcal{A}_{ij (11)}\ne\varnothing)\wedge (\mathcal{A}_{ij (22)}\ne\varnothing) \wedge (\mathcal{A}_{ij (12)}=\varnothing) \wedge (\mathcal{A}_{ij (21)}=\varnothing)]
\]
\begin{equation}
\label{eq:ar1.7}
\vee [(\mathcal{A}_{ij (11)}=\varnothing)\wedge (\mathcal{A}_{ij (22)}=\varnothing) \wedge (\mathcal{A}_{ij (12)}\ne\varnothing) \wedge (\mathcal{A}_{ij (21)}\ne\varnothing)] \textrm{.}
\end{equation}
Let's describe this selection rule by an operator $\hat A$. By definition, put $\hat A(\mathcal{B})=1$ if $\mathcal{B}\ne\varnothing$, were $\mathcal{B}$ is any set of chronons, and $\hat A(\varnothing)=0$. Let the number $L_{ij}$ be given by
\[
L_{ij}= \hat A(\mathcal{A}_{ij (11)}) \hat A(\mathcal{A}_{ij (22)}) (1-\hat A(\mathcal{A}_{ij (12)})) (1-\hat A(\mathcal{A}_{ij (21)}))
\]
\begin{equation}
\label{eq:ar1.8}
 +(1-\hat A(\mathcal{A}_{ij (11)})) (1-\hat A(\mathcal{A}_{ij (22)})) \hat A(\mathcal{A}_{ij (12)}) \hat A(\mathcal{A}_{ij (21)})\textrm{.}
\end{equation}
$L_{ij}=1$ if $\mathcal{A}_{ij}$ is a loop: $\mathcal{A}_{ij}=\mathcal{L}_{ij}$. Otherwise $L_{ij}=0$. Using (\ref{eq:es1.1}), (\ref{eq:ar1.4}), and (\ref{eq:ar1.8}), we get
\begin{equation}
\label{eq:ar1.9} p(\mathcal{L}_{ij})=p(\mathcal{A}_{ij})=\exp(-\ln2\langle \mathcal{A}_{ij} | L_{ij}\hat N | \mathcal{A}_{ij}\rangle) \textrm{.}
\end{equation}
Finally, we obtain
\begin{equation}
\label{eq:ar1.10} p(\mathcal{L}_{ij})=\exp(-\ln2\langle \mathcal{L}_{ij} | \hat N | \mathcal{L}_{ij}\rangle) \textrm{.}
\end{equation}
\subsection{State vectors of external monads \label{AR2}}
Consider a state vector $| \mathcal{X}_{is}\rangle$ of any x-structure $\mathcal{X}_i\equiv \{\alpha_{i1}, \alpha_{i2}, \beta_{i1}, \beta_{i2}\}$. The index $s$ ranges from 1 to 4. We can represent $| \mathcal{X}_{is}\rangle$ by a $1 \times 4$ column matrix. By $x_s$ denote the element of this matrix. By definition, put $x_1=1$ if $\alpha_{i1}$ is a maximal monad, otherwise $x_1=0$, $x_2=1$ if $\alpha_{i2}$ is a maximal monad, otherwise $x_2=0$, $x_3=1$ if $\beta_{i1}$ is a minimal monad, otherwise $x_3=0$, and $x_4=1$ if $\beta_{i2}$ is a minimal monad, otherwise $x_4=0$. This state vector is called the state vector of external monads of $\mathcal{X}_i$.

Let's assign a creation operator $\hat b^{\dag}_{is}$ to each external monad $\gamma_{is}$. The elements of $| \mathcal{X}_{is}\rangle$ can be only 0 or 1. We can assume that $\hat b^{\dag}$ is a Fermi creation operator.

Consider a d-graph $\mathcal{G}$ and two x-structures $\mathcal{X}_i\equiv \{\alpha_{i1}, \alpha_{i2}, \beta_{i1}, \beta_{i2}\}\in\mathcal{G}$ and $\mathcal{X}_j\equiv \{\alpha_{j1}, \alpha_{j2}, \beta_{j1}, \beta_{j2}\}\in\mathcal{G}$. Let $k$ be the number of the ordered pairs of external monads such that one external monad of the pair is included in $\mathcal{X}_i$ and other external monad of this pair is included in $\mathcal{X}_j$. Such pairs can exist only if $\alpha_{j1}\succ \beta_{i1}$ or $\alpha_{i1}\succ \beta_{j1}$. This is a selection rule. We can describe it by the number $ C_{ij}$. By definition, put $C_{ji}=1$, if $\beta_{i1}\prec\alpha_{j1}$, otherwise $C_{ji}=0$. We have
\begin{equation}
\label{eq:ar2.1}
k= \sum_{s=1}^4\sum_{r=1}^4\langle \mathcal{X}_{js} | \hat D | \mathcal{X}_{ir}\rangle \textrm{,}
\end{equation}
where
\begin{equation}
\label{eq:ar2.2}
\begin{array}{cccc}
\hat D &=&\left( \begin{array}{cccc} 0&0&C_{ji} \hat b^{\dag}_{j1}\hat b_{i3}& C_{ji} \hat b^{\dag}_{j1}\hat b_{i4}\\
\\0&0& C_{ji} \hat b^{\dag}_{j2}\hat b_{i3}& C_{ji} \hat b^{\dag}_{j2}\hat b_{i4}\\
\\ C_{ij} \hat b^{\dag}_{j3}\hat b_{i1}& C_{ij} \hat b^{\dag}_{j3}\hat b_{i2}&0&0\\
\\ C_{ij} \hat b^{\dag}_{j4}\hat b_{i1}& C_{ij} \hat b^{\dag}_{j4}\hat b_{i1}&0&0 \end{array} \right) &\textrm{.}
\end{array}
\end{equation}
Using (\ref{eq:es4.1}), we get
\begin{equation}
\begin{array}{l}
\label{eq:ar2.3} p(\mathcal{S})=\exp[(\ln2)q \sum_{s=1}^4\sum_{r=1}^4\langle \mathcal{X}_{js} | \hat D | \mathcal{X}_{ir}\rangle] \textrm{,}
\end{array}
\end{equation}
where $\mathcal{S}$ is the pair of x-structures $\mathcal{X}_i$ and $\mathcal{X}_j$.
\subsection{The equation of motion \label{AR3}}
Consider the elementary extension to the future of $\mathcal{G}$. This is the addition of the new x-strucrure $\mathcal{X}_j$. The probability of this elementary extension depends on all dynamical structures that are generated by $\mathcal{X}_j$. Using (\ref{eq:ar1.9}) and (\ref{eq:ar2.3}), we can write the equation of motion (\ref{eq:es4.3}) in the algebraic representation.
\[
P=C\, exp[(\ln2) \sum_{i=1}^N (q \sum_{s=1}^4\sum_{r=1}^4\langle \mathcal{X}_{js} | \hat D | \mathcal{X}_{ir}\rangle
\]
\begin{equation}
\label{eq:ar3.1}
-\langle \mathcal{A}_{ij} | L_{ij}\hat N | \mathcal{A}_{ij}\rangle)]\textrm{,}
\end{equation}
where $N$ is the number of x-structures in $\mathcal{G}$.

Similarly, consider the elementary extension to the past of $\mathcal{G}$. This is the addition of the new x-strucrure $\mathcal{X}_i$. We have
\[
P=C\, exp[(\ln2) \sum_{j=1}^N (q \sum_{s=1}^4\sum_{r=1}^4\langle \mathcal{X}_{js} | \hat D | \mathcal{X}_{ir}\rangle
\]
\begin{equation}
\label{eq:ar3.2}
-\langle \mathcal{A}_{ij} | L_{ij}\hat N | \mathcal{A}_{ij}\rangle)]\textrm{.}
\end{equation}

The dynamical law is not a differential equation but $P$ constructed by finite algebraic operations. If multipliers in the equation of motion depend on the number of chronons in some sets this is a discrete protoform of a Bose field. In the considered example this is the loop multipliers. The causal connection can have only two values: ``the causal connection exists'' and ``the causal connection does not exist''. If multipliers in the equation of motion depend on the causal connection of monads this is a discrete protoform of a Fermi field. In the considered example this is the multipliers of ordered pairs of external monads. Each x-structure has 4 monads. Consequently there are 4 components in a state vector of external monads in an x-structure. This is a discrete protoform of a spinor field. In the considered example the discrete protoform of a spinor field coincides with a discrete protoform of a Fermi field.
\section{CONCLUSION \label{CON}}
This model illustrates how matter can arise dynamically from a discrete pregeometry without having to be built in at the fundamental level. At the root of the causal set program is the recognition that, when it is combined with volume information, causal structure alone suffices to reproduce fully the geometry of spacetime. In brief ``Order + Number = Geometry''. The considered program develops this idea. The d-graph suffices to reproduce fully the reality. In brief ``Order + Number = Matter + Geometry''.

An open important question, of course, is an emergent spacetime. Is there a procedure by which, given a d-graph, one can determine whether there is a manifold that approximates it, and possibly construct such a manifold if one exists? Spacetime is only an emergent manifestation of a matter structure on a larger scale. Firstly, we must construct a matter. Secondly we must construct a net of intersecting world lines spanning the spacetime continuum.

Our first goal will be to describe particles. What structures of a d-graph correspond to each particle? We must describe the properties of quantum particles in terms of a d-graph to answer this question. According to a d-graph approach, the world is a collection of processes with a causal structure. Then an object is secondary; is a long causal sequence of processes, a world line. According to quantum mechanics the world is a collection of objects (particles). Then a process is secondary; is a mapping of the objects or of their initial to their final conditions. The translation from the language of quantum mechanics is not obvious. We need the rules of interpretation that assign particular quantum numbers to particular properties of d-graph. Perhaps, the further development of the algebraic description of a d-graph will be useful.

Also, the open question is the approach to the evolution of the universe. The considered model can describe only several interacting particles. We might continue multiplying questions, but let's finish.
\section*{ACKNOWLEDGEMENTS}
I am grateful to several colleagues for extensive discussions on this subject, especially Alexandr V. Kaganov and Vladimir V. Kassandrov.

\end{document}